\documentclass[fleqn,usenatbib]{mnras}

\usepackage{newtxtext,newtxmath}

\usepackage[T1]{fontenc}

\DeclareRobustCommand{\VAN}[3]{#2}
\let\VANthebibliography\thebibliography
\def\thebibliography{\DeclareRobustCommand{\VAN}[3]{##3}\VANthebibliography}

\usepackage{graphicx}	
\usepackage{amsmath}	

\usepackage{amssymb}	
\usepackage{siunitx} 
\usepackage{booktabs}
\usepackage[flushleft]{threeparttable}
\usepackage{float}
\usepackage{cleveref}
\usepackage{pifont}
\usepackage{hyperref}

\input{anc/epcommands}
\newcommand{\PHOEBE}{\texttt{PHOEBE}}

\defcitealias{Rowan22}{R22}
\defcitealias{Zasche22}{Z22}
\defcitealias{Kostov22}{K22}
\defcitealias{Rappaport22}{R22}
\defcitealias{Zasche19}{Z19}

\title[Extra-Physics Binaries]{The Value-Added Catalog of ASAS-SN Eclipsing Binaries II: Properties of Extra-Physics Systems}

\author[D. M. Rowan et al.]{D. M.
Rowan,$^{1,2}$\thanks{E-mail: rowan.90@osu.edu},
T. Jayasinghe$^{1,2,3,4}$,
K. Z. Stanek$^{1,2}$,
C. S. Kochanek$^{1,2}$,
Todd A. Thompson$^{1,2,5}$,
\newauthor
B. J. Shappee$^{6}$,
T. W. -S. Holoien$^{7}$,
J. L. Prieto$^{8,9}$, and
W. Giles$^{10}$
\\
$^{1}$Department of Astronomy, The Ohio State University, 140 West 18th Avenue, Columbus, OH, 43210, USA\\
$^{2}$Center for Cosmology and Astroparticle Physics, The Ohio State University, 191 W. Woodruff Avenue, Columbus, OH, 43210, USA\\
$^{3}$Department of Astronomy,  University of California Berkeley, Berkeley CA 94720, USA\\
$^{4}$NASA Hubble Fellow\\
$^{5}$Department of Physics, The Ohio State University, Columbus, Ohio, 43210, USA\\
$^{6}$Institute for Astronomy, University of Hawaii, 2680 Woodlawn Drive, Honolulu, HI 96822, USA\\
$^{7}$Carnegie Observatories, 813 Santa Barbara Street, Pasadena, CA 91101, USA\\
$^{8}$N\'ucleo de Astronom\'ia de la Facultad de Ingenier\'ia y Ciencias, Universidad Diego Portales, Av. Ej\'ercito 441, Santiago, Chile\\
$^{9}$Millennium Institute of Astrophysics, Santiago, Chile\\
$^{10}$ASC Technology Services, 433 Mendenhall Laboratory 125 South Oval Mall Columbus OH, 43210, USA\\
}
\date{Accepted XXX. Received YYY; in original form ZZZ}

\pubyear{2022}

\begin{document}
\label{firstpage}
\pagerange{\pageref{firstpage}--\pageref{lastpage}}
\maketitle

\begin{abstract}
Detached eclipsing binaries are the primary tool used to measure precise masses and radii of stars. In our previous paper estimating the parameters of more than 30,000 detached eclipsing binaries, we identified \nEP{} eclipsing binaries with additional features in their All-Sky Automated Survey for Supernovae (ASAS-SN) and Transiting Exoplanet Survey Satellite (TESS) light curves. Here, we characterize these ``extra-physics'' systems, identifying eclipsing binaries with spotted stars, pulsating components, and candidate triple/quadruple systems. We use the Gaia, ATLAS, ZTF, and ASAS-SN variable star catalogs to consider possible blends. We use MIST isochrones and evolutionary tracks to identify systems with main sequence, subgiant, and giant primaries and highlight systems in sparsely populated regions of the color-magnitude diagram. We find that the orbital period distribution of spotted binaries is divided by evolutionary state and find \nXraySpotted{} with X-ray detections. For the candidate triple/quadruples and pulsating systems, we calculate the extra orbital/pulsational period and identify systems with resonances. Finally, we highlight a number of exotic systems, including eclipsing CVs, subdwarfs, and binaries with disks. 
\end{abstract}




\section{Introduction}

One of the primary applications of eclipsing binaries is to determine the physical parameters of stars. By modeling their light curves and radial velocities, precise masses and radii can be determined and used to validate and improve models of stellar structure \citep[e.g.,][]{Andersen91, Torres10}.  Photomometric and spectroscopic surveys have discovered tens of thousands of binary stars \citep{Graczyk11, Prsa11, Petrosky21, Pojmanski02, JayasingheII, Jayasinghe21, Christy22, Pourbaix04}. Interesting subsets of systems can then be identified from such large catalogs, such as extremely eccentric binaries \citep[e.g.,][]{Zasche21}, systems with changing eclipse depths \citep[e.g.,][]{Davenport21}, and triple and quadruple eclipsing systems \citep[e.g.,][]{Rappaport22, Kostov22, Zasche22}.

Eclipsing binaries are traditionally classified into four morphological classes. Detached eclipsing binaries or Algol-type binaries have essentially spherical stars with no obvious effects of ellipsoidal variability in the light curve. Semidetached systems ($\beta$-Lyrae) have one or both stars nearly filling their Roche lobes, and contact binaries (W Uma) fill their Roche lobes. Finally, ellipsoidal variables are binaries where the modulations from tidal distortion dominate the overall light curve shape, often without observed eclipses \citep[e.g.,][]{Rowan21}. The majority of eclipsing binary catalogs use these four classifications \citep[e.g.,][]{JayasingheII}, although some catalogs instead use a continuous variable to describe the `detachedness' of the binary \citep{Matijevic12, Prsa22}.

Subclasses of eclipsing binaries have also been introduced for eclipsing binaries with additional light curve features. For example, the RS Canum Venaticorum-type binary systems (RS CVn) have increased stellar activity and spots producing sinusoidal features in the light curve that change over time. The identification and modeling of eclipsing RS CVn systems can be used to improve our understanding of stellar activity cycles and the evolutionary history of the binary \citep{Roettenbacher16}. 

Pulsational variability of one or both of the components in an eclipsing binary can be used to study stellar structure in greater detail \citep{Kahraman17}. Properties of the interior structure of the star can be determined by modeling the pulsation characteristics, and the physical mass and radius can be determined by modeling the light curve and radial velocity curve. These systems can also be used to understand how pulsations are affected by the gravitational force of the binary companion \citep{Soydugan06a}. Eclipsing binaries containing various types of pulsators have been identified, including $\delta$~Scuti \citep{Soydugan06b, Shi22}, $\gamma$~Dor \citep{Damiani10}, and Cepheid \citep{Gieren15} variables.

Eclipsing binary catalogs have also aided in the search for systems with a third body or doubly-eclipsing binaries. The detection and analysis of these systems can be used to improve our understanding of the frequency of higher-order systems, their formation mechanisms, and the dynamical processes that shape their evolution. Tertiary companions can be identified in binary systems by monitoring the eclipse timing variations over timescales much longer than the orbital period \citep{Borkovits15}. Depending on the orbital inclination of the third body, they can also be discovered by detecting additional eclipses \citep{Marsh14, Rappaport22}. Similarly, quadruple systems made up of two pairs of eclipsing binaries can be found by searching for additional periods in the light curves of known eclipsing binaries \citep{Kostov22, Zasche22}. Catalogs of doubly eclipsing binaries suggest an excess of systems with 3:2 period ratios \citep{Zasche19} which can be used to inform models of resonant capture \citep{Tremaine20}.

Machine learning techniques are almost always used to classify variable stars, including eclipsing binaries. However, the detection of unusual subclasses of systems, like pulsating eclipsing binaries or doubly eclipsing binaries, is still often done through visual inspection. In \citet[][hereafter R22]{Rowan22}, we visually inspected more than 40,000 eclipsing binary light curves from the All-Sky Automated Survey for Supernovae \citep[ASAS-SN,][]{Shappee14, Kochanek17} and Transiting Exoplanet Survey Satellite \citep[TESS,][]{Ricker15, Kunimoto21, Huang20a, Huang20b} to verify our eclipsing binary models. In the process of we also identified \nEP{} systems that could not be be described by simple models of eclipsing binaries due to the presence of spots, pulsations, or additional eclipses. 

Here we present the ASAS-SN $V$, ASAS-SN $g$, and TESS $T$ band light curves and discuss the properties of these \nEP{} eclipsing binaries with extra physics. In Section \S\ref{sec:lc_identification}, we describe the ASAS-SN and TESS light curves and the visual inspection process used to identify and classify these light curves. In Section \S\ref{sec:catalog_description}, we show the distribution of light curve parameters, including the Gaia DR3 CMD position. Sections \S\ref{sec:spotted_stars}--\ref{sec:gamma_dor_pulsators}, we describe the properties of the different variability classes and calculate the periods of pulsations, triples, and quadruple systems when applicable. Finally, we characterize unique systems that do not fall into the groups discussed in Section \S\ref{sec:lc_identification} in Section \S\ref{sec:other_targets}. 

\section{Light curves and target identification} \label{sec:lc_identification}

In \citetalias{Rowan22}, we modeled the ASAS-SN $V$- and $g$-band light curves of more than 30,000 detached eclipsing binaries from \citet{Jayasinghe21}. The $V$-band light curves span from 2012 to mid 2018 and the $g$-band observations have been ongoing since late 2017. The optimal magnitude range for ASAS-SN targets is $11 < V < 17$~mag and $12 < g < 18$~mag \citep{JayasingheII}. 

We used \PHOEBE{} \citep{Prsa05, Prsa16, Conroy20}, a tool for modeling eclipsing binaries of all morphological classes, to fit the ASAS-SN light curves, and then we visually inspected all the model fits to identify systems that required additional optimization. As a part of this process, we downloaded the TESS light curves from the the SPOC \citep{Caldwell20} and QLP \citep{Huang20a, Huang20b} pipelines. We use the ``raw'' light curves rather than the detrended light curves since the detrending procedure can often remove real stellar variability. We folded the TESS light curves at the period calculated from the ASAS-SN light curves and identified \nEP{} targets that showed additional features in the ASAS-SN or TESS light curves. At the time of writing (October 2022), TESS light curves are available for \nMastTESS{} of our targets.

We broadly group these systems by the nature of the ``extra physics'' seen in the light curve: 
\begin{itemize}
    \item Spotted stars: \nSpotted{} targets
    \item Triple and doubly eclipsing binaries: \nEE{} targets
    \item Reflection effects: \nREFL{} targets
    \item Regular pulsators: \nRP{} targets 
    \item Stochastic variability: \nSto{} targets
    \item $\gamma$-Dor pulsators: \nIRP{} targets
    \item Other: \nOther{} targets
\end{itemize}

\begin{figure*}
    \centering
    \includegraphics[width=\linewidth]{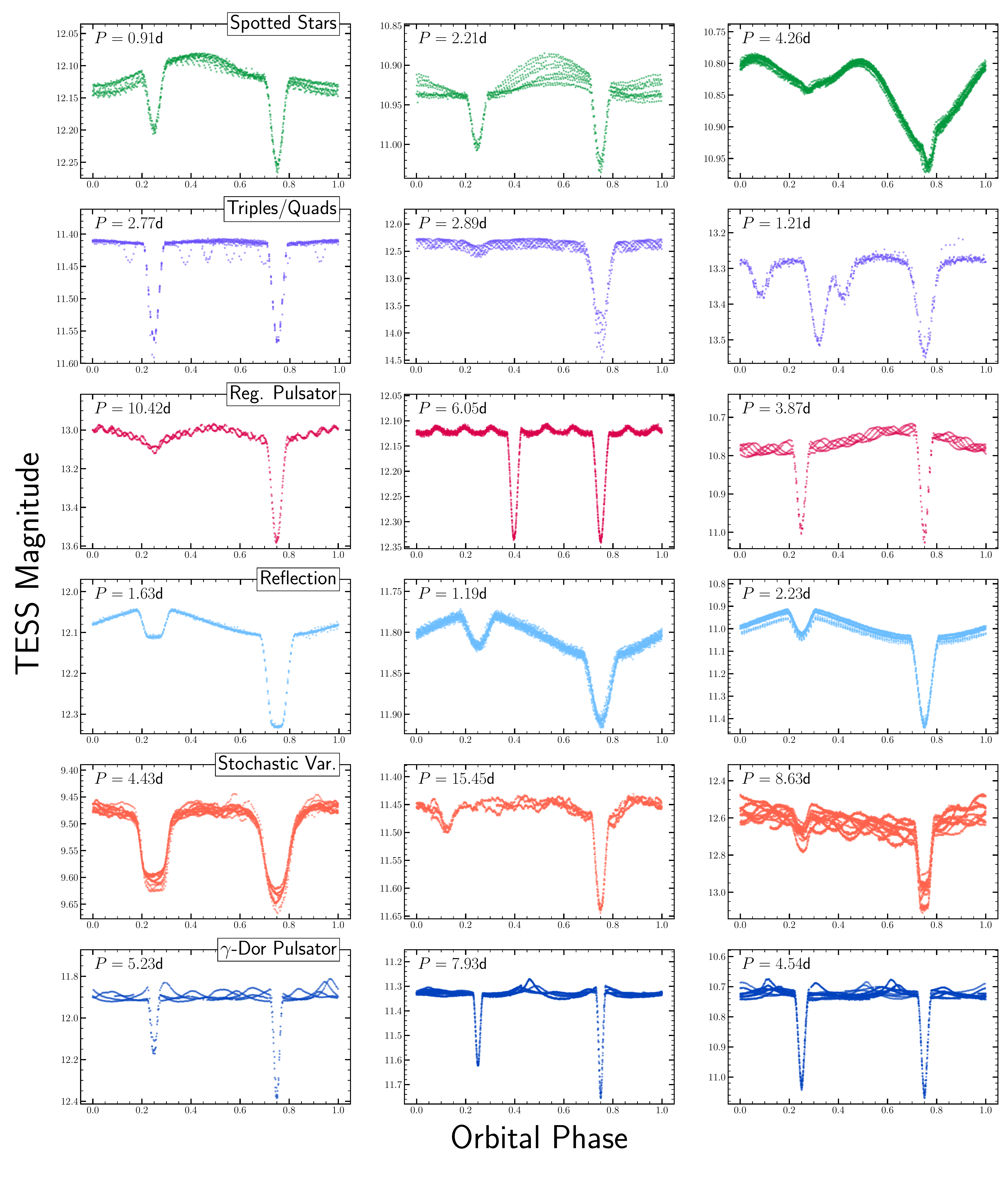}
    \vspace{-1cm}
    \caption{TESS light curves for examples of the different variability types. The orbital period is labeled in the upper left of each panel. The light curves are folded so that the primary eclipse occurs at phase 0.75.}
    \label{fig:panel_plot}
\end{figure*}

Figure \ref{fig:panel_plot} shows the TESS light curves for example systems from each group. The largest group are the spotted systems (\nSpotted{} targets) where one or more spots distort the overall shape of the light curve. In some cases the amplitude and phasing of the spots changes over the course of the TESS sector or between sectors. The next largest group are the systems with extra eclipses (\nEE{} targets) from a tertiary or an additional eclipsing binary. In the phased light curve, these systems are identified as having additional eclipses at a different orbital period. We also identify \nRP{} systems showing regular, periodic pulsations, as well as \nIRP{} binaries with more $\gamma$-Dor-like variability and \nSto{} systems with stochastic variability. Finally, we identify \nREFL{} systems where the large temperature differences between the components results in an overall brightening of the light curve near the secondary eclipse due to irradiation. There are also \nOther{} targets that do not fit into any of these categories. These are discussed in Section \ref{sec:other_targets}. The full set of light curves are available online at \url{https://asas-sn.osu.edu/binaries}. Table \ref{tab:ep_table} gives the parameters of each system and their classification into the groups listed above. 

\begin{table*}
    \centering
    \caption{Parameters of the extra-physics sample. The $\texttt{r1}$ statistic comes from the ATLAS REFCAT-2 catalog and is set at 99.9 if the cumulative flux from nearby stars does not equal the flux of the target within $36\farcs0$. The Gaia absolute magnitude ($M_G$) and color ($G_{\rm{BP}}-G_{\rm{RP}}$) are corrected for extinction using {\tt mwdust}. The evolutionary state is calculated based on the procedure described in \citetalias{Rowan22}. $P_2$ gives the period of the pulsations or additional eclipses. The reference column gives the citation for triple/quadruple systems included in existing catalogs. The full table is available online at \url{https://asas-sn.osu.edu/binaries} and in the electronic version of the paper.}
    \sisetup{table-auto-round,
         group-digits=false}
    \setlength{\tabcolsep}{3pt}
    \begin{center}
        \begin{tabular}{l S[table-format=3.4] l S[table-format=2.1] S[table-format=2.2] S[table-format=2.2] r S[table-format=3.4] r r S[table-format=2.1] r}
\toprule
  {Gaia DR3 Source} &  {Period} &               {Group} &  {$\texttt{r1}$} &   {$M_G$} &  {$G_{\rm{BP}}-G_{\rm{RP}}$} & {State} &    {$P_2$} &                                 {Reference} &              {$L_X$} & {X-ray Sep.} & {X-ray Obs.} \\
{} & {(d)} & {} & {(")} & {(mag)} & {(mag)} & {} & {(d)} & {} & {(erg/s)} & {(")} & {}\\ 
\midrule
4634512158994726144 &  6.038744 &         Spotted Stars &             99.9 &  5.744441 &                     1.269236 &         &   6.031107 &                                             &                      &              &              \\
2880801592412343552 & 16.031115 &         Spotted Stars &             99.9 &  3.049203 &                     1.100450 &      SG &  16.090629 &                                             &                      &              &              \\
4991376490493277952 &  0.950549 &         Spotted Stars &             99.9 &  4.920895 &                     0.886524 &      MS &   0.947022 &                                             & $1.3 \times 10^{31}$ &     3.935197 &        SWIFT \\
 430569204423511552 &  0.848522 &         Triples/Quads &             99.9 &  3.278892 &                     0.634677 &      MS &   3.517713 &                                             &                      &              &              \\
 431081611195489536 &  2.354659 &         Triples/Quads &             99.9 & -2.351562 &                     0.116708 &      SG &            &                                             &                      &              &              \\
 524420699465597184 &  4.079206 &         Triples/Quads &             99.9 & -1.395212 &                     0.213919 &      SG &     4.9867 & \citetalias{Zasche22},\citetalias{Kostov22} &                      &              &              \\
 401523955495860608 & 16.738921 &         Reg. Pulsator &             99.9 &  1.710154 &                     0.541307 &      MS &   0.951298 &                                             &                      &              &              \\
 457177145291803776 &  3.489172 &         Reg. Pulsator &             99.9 & -0.780096 &                     0.045592 &         &   0.530743 &                                             &                      &              &              \\
 457720785072612480 &  4.606832 &         Reg. Pulsator &             99.9 & -0.103586 &                     0.229628 &      MS &   1.057468 &                                             &                      &              &              \\
 513586146241283072 &  1.631960 &            Reflection &             99.9 & -1.005741 &                     0.367331 &         &            &                                             &                      &     0.267248 &      CHANDRA \\
 465551472526065152 &  1.321082 &            Reflection &             99.9 & -1.782264 &                    -0.038696 &      MS &            &                                             & $1.5 \times 10^{31}$ &     1.064106 &   XMM-NEWTON \\
3449142279545037056 &  1.619217 &            Reflection &             99.9 & -1.302655 &                     0.040003 &      MS &            &                                             &                      &              &              \\
 511009646904766336 & 11.722420 &       Stochastic Var. &             99.9 & -3.962492 &                     0.081498 &         &            &                                             &                      &              &              \\
 461470115422532352 &  2.127926 &       Stochastic Var. &             99.9 & -1.979391 &                    -0.044067 &      SG &            &                                             &                      &              &              \\
 491478678258597632 &  3.139911 &       Stochastic Var. &             99.9 &  2.434521 &                     0.788118 &      SG &            &                                             &                      &              &              \\
 351850661555541504 &  5.226429 & $\gamma$-Dor Pulsator &             99.9 &  2.689258 &                     0.578219 &      MS &            &                                             &                      &              &              \\
  23092256739885312 &  5.681869 & $\gamma$-Dor Pulsator &             99.9 &  1.979730 &                     0.477387 &      MS &            &                                             &                      &              &              \\
 435896716209335040 &  8.293102 & $\gamma$-Dor Pulsator &             99.9 &  2.149057 &                     0.505999 &      MS &            &                                             &                      &              &              \\
4687138015328580736 & 15.081228 &                 Other &             99.9 & -2.360697 &                    -0.117241 &         &            &                                             &                      &              &              \\
4655260836825948160 & 27.111238 &                 Other &             21.9 & -2.868630 &                    -0.665235 &         &            &                                             &                      &              &              \\
6055907396399148032 &  2.533139 &                 Other &             99.9 &  1.605868 &                     0.631236 &      MS &            &                                             &                      &              &              \\
\bottomrule
\end{tabular}

    \end{center}
    \label{tab:ep_table}
\end{table*}

\section{Catalog description} \label{sec:catalog_description}

\begin{figure*}
    \centering
    \includegraphics[width=\linewidth]{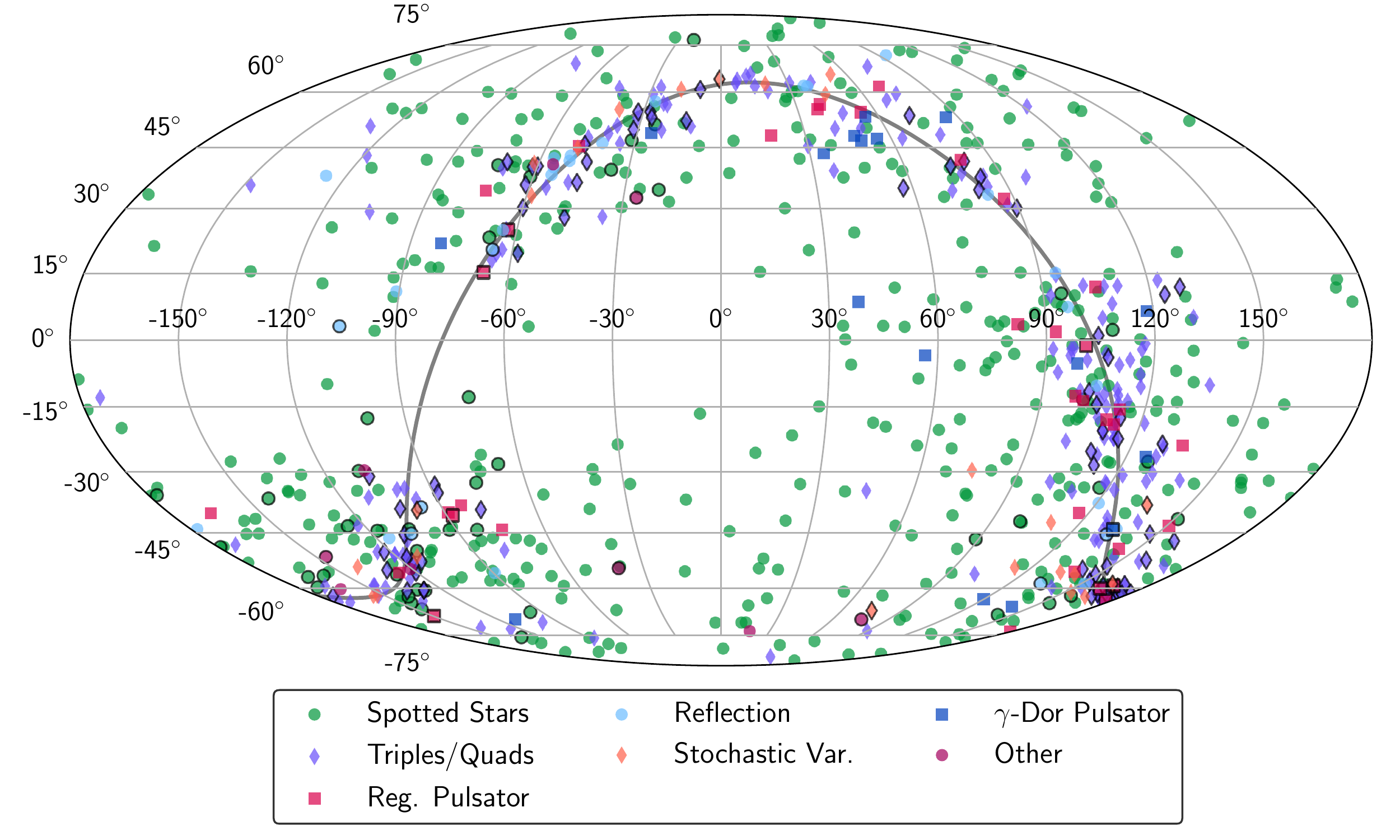}
    \caption{Sky distribution of the extra-physics targets in equatorial coordinates. The Galactic plane is the gray line. Targets are colored by their variability type. Targets with $\texttt{r1}<36\arcsec$ are shown with black borders. We find an excess of targets near the Galactic plane with $\texttt{r1}<36\arcsec$, especially for systems showing extra eclipses, suggesting some could be blends due to the low resolution of TESS.}
    \label{fig:skymap}
\end{figure*}

\begin{table*}
    \centering
    \caption{Cross matches of the extra-physics targets to the Gaia, ZTF, and ATLAS variability catalogs with a radius of $21\arcsec$. The Gaia Source and magnitude from Table \ref{tab:ep_table} are given in the first two columns. The $P_2$ column reports the Gaia, ZTF, or ATLAS period if it is within 5\% the rotation period, spotted period, or second orbital period in the TESS light curve. The full table is available online at \url{https://asas-sn.osu.edu/binaries} and in the electronic version of the paper.}
    \sisetup{table-auto-round,
         group-digits=false}
    \setlength{\tabcolsep}{1.5pt}
    \begin{center}
        \begin{tabular}{l S[table-format=2.1] | l S[table-format=2.1] S[table-format=2.1] l S[table-format=2.1] S[table-format=2.1] l S[table-format=2.1] S[table-format=2.1] S[table-format=2.3]}
\toprule
      {DR3 Source} &     {$G$} & {Nearby DR3 Source} &      {$G$} & {Gaia Sep.} &            {ZTF ID} &   {$g$} & {ZTF Sep.} &        {ATLAS ID} &   {$o$} & {ATLAS Sep.} &    {$P_2$} \\
{} & {(mag)} & {} & {(mag)} & {(")} & {} & {(mag)} & {(")} & {} & {(mag)} & {(")} & {(d)}\\ 
\midrule
183120431790693120 & 12.875074 &                     &            &             &                     &         &            & J083.5330+35.2988 &  16.246 &    14.028555 &            \\
198407697866912128 & 11.543592 &                     &            &             &                     &         &            & J073.9465+37.3475 &  16.061 &    20.233713 &            \\
200791134137604352 & 13.760879 &                     &            &             &                     &         &            & J074.9785+40.4087 &  13.605 &      0.02314 &   3.786638 \\
219832339066395264 & 11.622397 &  219832339066395392 &  12.143232 &    2.468729 &                     &         &            &                   &         &              &            \\
239043654764555520 & 13.054091 &                     &            &             &                     &         &            & J046.7897+39.9546 &  15.644 &    11.342994 &            \\
273109278071250944 & 13.792052 &                     &            &             & J044752.88+533026.5 &  15.291 &   7.759133 &                   &         &              &            \\
337798490636307968 & 12.628793 &                     &            &             & J022611.88+391415.8 &  14.726 &   8.176808 &                   &         &              &  13.685183 \\
355800111620808576 & 12.517892 &                     &            &             & J020003.55+452605.2 &  12.992 &   6.144629 & J030.0148+45.4347 &  12.715 &     6.107379 &    0.58915 \\
434846854403757184 & 13.920694 &                     &            &             &                     &         &            & J049.0648+46.7908 &  15.445 &     8.089793 &            \\
434846854403757184 & 13.920694 &                     &            &             &                     &         &            & J049.0645+46.7950 &  15.849 &    13.527997 &            \\
447114998986540672 & 12.008864 &                     &            &             &                     &         &            & J046.9632+54.0670 &  14.265 &     8.756023 &            \\
447114998986540672 & 12.008864 &                     &            &             &                     &         &            & J046.9594+54.0661 &  11.911 &     0.001915 &   2.407308 \\
505860530871568512 & 13.355546 &  505860526571955712 &  19.554836 &    8.769168 & J014430.68+572555.5 &  20.444 &   8.842957 &                   &         &              &            \\
510360419650993920 & 13.041862 &                     &            &             & J012033.99+602257.2 &  18.002 &  19.762312 &                   &         &              &            \\
524381426285320064 & 13.464093 &                     &            &             & J005638.57+652706.8 &   17.75 &   9.379235 &                   &         &              &            \\
\bottomrule
\end{tabular}

    \end{center}
    \label{tab:contamination_table}
\end{table*}

\begin{figure}
    \centering
    \includegraphics[width=\linewidth]{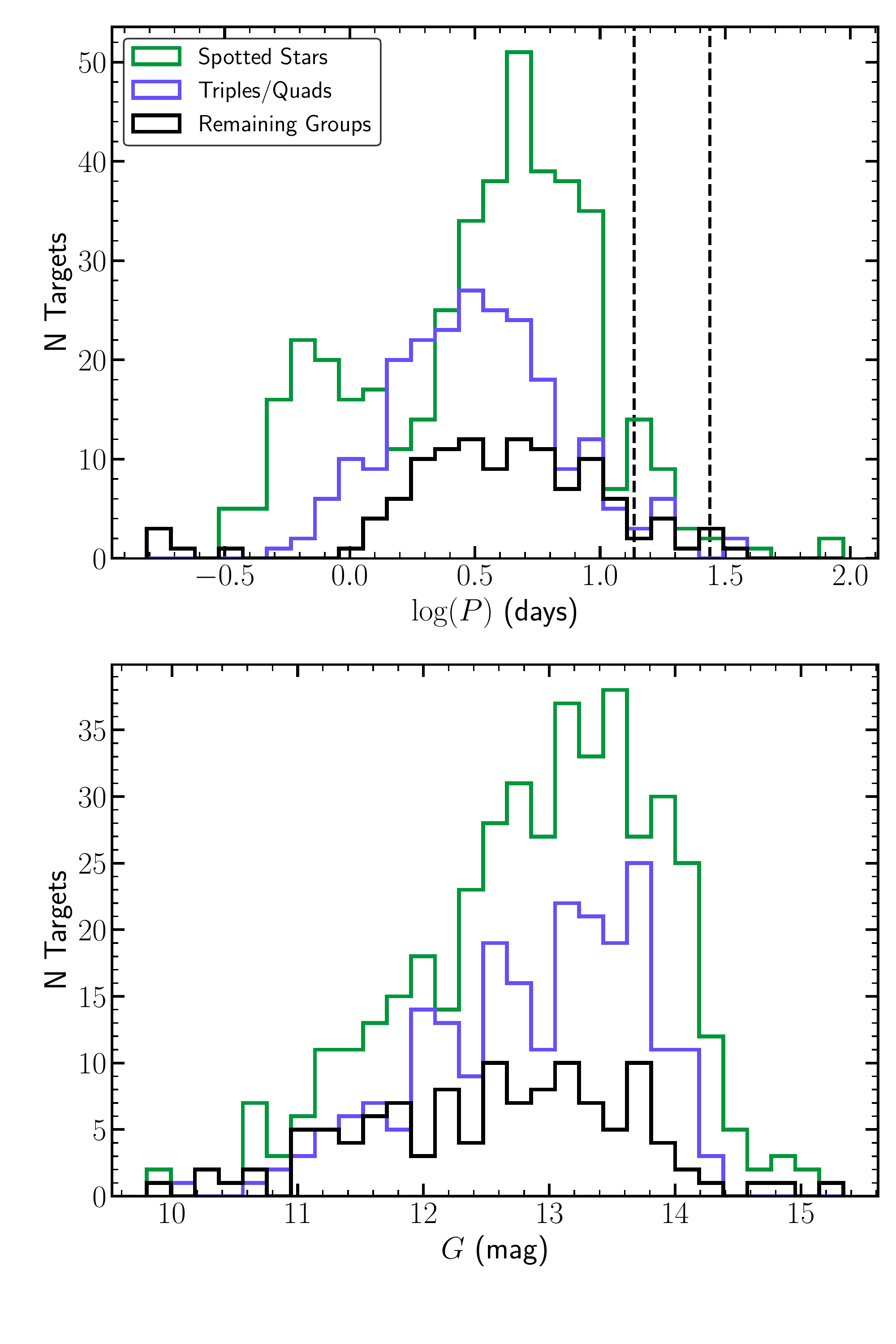}
    \caption{Top: distribution of orbital periods. Even though these targets were originally selected from the multi-year ASAS-SN data, we find that the majority of targets have periods shorter than the TESS orbit/sector length, shown by the vertical dashed lines. Bottom: the distribution of Gaia $G$ magnitudes.}
    \label{fig:param_hists}
\end{figure}

\begin{figure*}
    \centering
    \includegraphics[width=\linewidth]{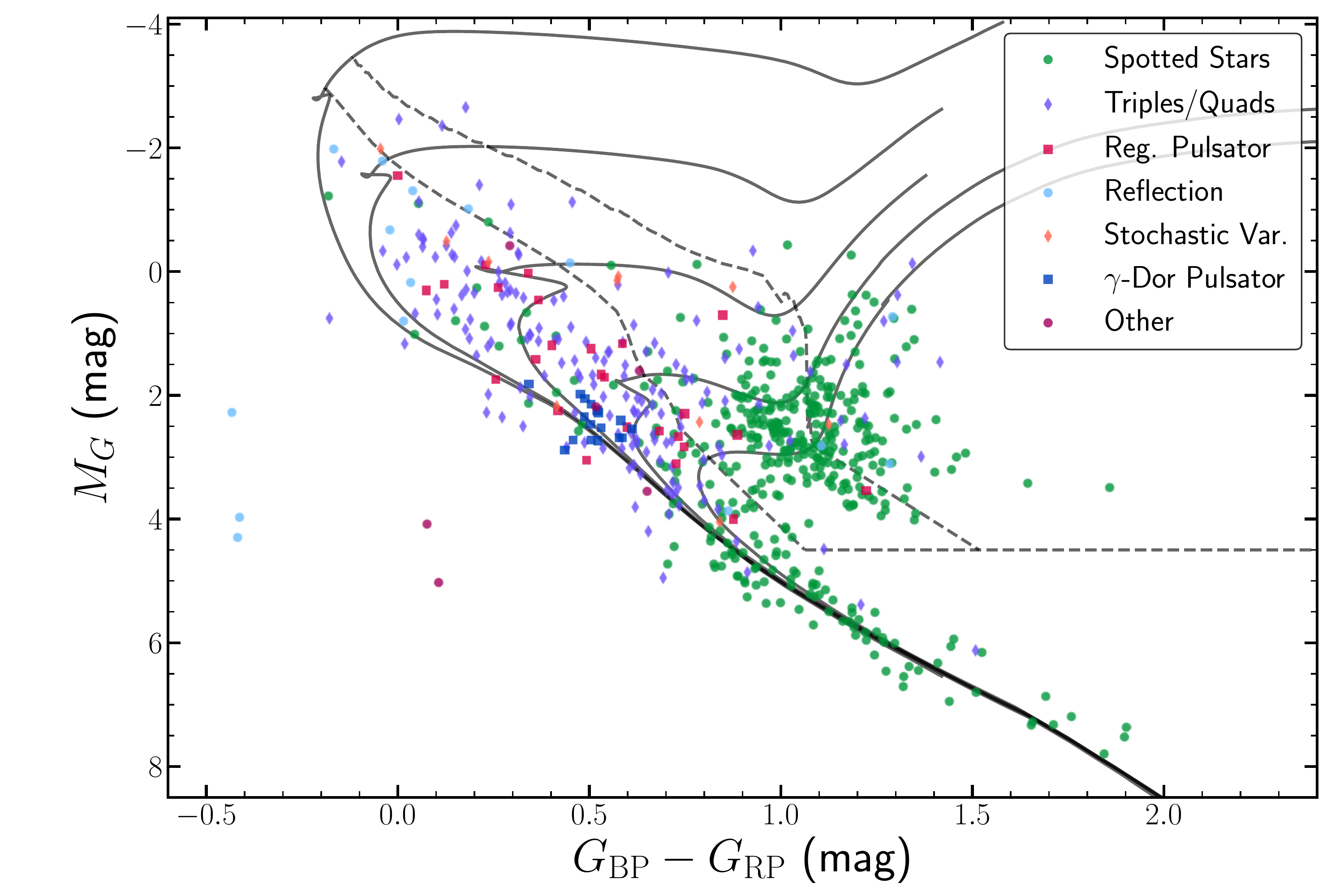}
    \caption{Gaia DR3 extinction-corrected color-magnitude diagram (CMD) colored by variability type. The solid lines show MIST isochrones for ages of $10^8$ to $10^{10}$ years in intervals of $0.5$ dex. The flux is doubled in each band to represent binary stars of equal mass. The dashed lines show the boundaries of the giant and subgiant branches defined by \citetalias{Rowan22}. The five systems are significantly below the main sequence are discussed in Sections \ref{sec:reflection_systems} and \ref{sec:other_targets}.}
    \label{fig:cmd}
\end{figure*}

Figure \ref{fig:skymap} shows the distribution of the systems on the sky. Since the TESS pixels are $21\arcsec$ and typical aperture radii range from 1.75 to 8 pixels \citep{Huang20a}, it is possible that some of these systems, especially those with extra eclipses, are actually blended sources.  To quantify how many of these targets may be blends, we cross-match with the ATLAS all-sky stellar reference catalog \citep[ATLAS REFCAT2, ][]{Tonry18}. The ATLAS {\tt r1} proximity statistic gives the radius where the cumulative flux of nearby stars equals the flux of the target up to a search radius of $36\arcsec$. Targets $\texttt{r1}<36\arcsec$ are more likely to be blends. Out of the \nEP{} extra-physics targets, \nEPrI{} have $\texttt{r1} < 36\arcsec$. In Figure \ref{fig:skymap} the targets with $\texttt{r1} < 36\arcsec$ are marked with black borders and we see these are concentrated in the Galactic plane.

If a target is actually a blend, we may expect to find multiple matches to variability catalogs for a single source. We cross-match our catalog with the Gaia DR3 \citep{GaiaCollab2022}, ZTF \citep{Chen20}, ATLAS \citep{Heinze18}, and ASAS-SN \citep{Jayasinghe21, Christy22} variability catalogs with a search radius of $21\arcsec$. All of these surveys have higher resolution than TESS, so blended targets are more likely to be identified as distinct variable sources. For the Gaia crossmatch, we take all targets within $21\arcsec$ that have the {\tt phot\_variable\_flag} set to ``variable''. The Gaia variability catalog may be less reliable for evaluating the blending of triple/quadruple systems, since detached eclipsing binaries are less likely to be identified in the sparse, sigma-clipped Gaia light curves. Out of the \nEP{} targets in our catalog, only \nVarFlag{} binaries (\FracVarFlag{}\%) are identified as variable in Gaia DR3. For the ZTF cross-match, we consider targets in the main variability catalog and the suspected variable catalog \citep[Tables 2 and 3 of ][]{Chen20}. Only 372 and 80 binaries are identified as variables in the ATLAS and ZTF catalogs, respectively, but we note that these catalogs are not all-sky.

We also compare the Gaia, ATLAS, and ZTF periods, when available, to the orbital period, $P$, and the second orbital period, $P_2$ (calculated as described below in Sections \S\ref{sec:spotted_stars}, \S\ref{sec:extra_eclipses} and \S\ref{sec:regular_pulsators}). We check if the periods given in the variability catalogs are within 5\% of $P$ or $P_2$, and also check for aliases at half and double these periods. With the ZTF catalog, we compare with the $g$- and $r$-band periods, and for ATLAS we compare with the periods from the long-period Fourier fit, short-period Fourier fit, and Lomb-Scargle periodogram. For the Gaia variables, we consider the periods corresponding to the fundamental and first overtone mode of variables in the RR Lyrae and Cepheid catalogs, the periods in the short- and long-timescale variability catalogs, the main sequence oscillator periods, rotation periods, and eclipsing binary periods.

To remove the variable star entry corresponding to the original ASAS-SN eclipsing binary in the Gaia cross-match, we simply check that the Gaia Source is different from the that of the original ASAS-SN source. For the ATLAS and ZTF cross-match, we assume that the variable star corresponds to the original ASAS-SN eclipsing binary if the ATLAS/ZTF period is consistent with $P$, or if it is separated by $<5\arcsec$. We also flag any variable with separation $<21\arcsec$ where the period is within 5\% of $P_2$ (or its aliases) as the highest confidence blended targets. 

Table \ref{tab:contamination_table} reports the available matches. In total, \nMatchBlends{} of our binaries are found to have nearby variable stars. The largest number of matches are found in Gaia DR3 (\nMatchBlendsGaia{} matches), followed by ATLAS (\nMatchBlendsATLAS{} matches), and ZTF (\nMatchBlendsZTF{} matches). We find no matches in the ASAS-SN $V$- and $g$-band catalogs. The triple/quadruple systems have the most number of possible blended variables (\nMatchBlendsEE{} binaries) followed by spotted stars (\nMatchBlendsSpotted{} binaries). The remaining groups have $<10$ systems with nearby variable stars. We find that \nMatchBlendsPII{} of these systems have periods potentially consistent with the period of the extra eclipses or the pulsation period. These systems have the most evidence for being blends in the TESS light curves, but we note that \nMatchBlendsPIIlowSep{} of these nearby variables are separated by $<5\arcsec$, which may indicate that the system is either a blend of two unresolved binaries in the ZTF or ATLAS light curves, or is truly multi-periodic, with the second period being reported in the ATLAS/ZTF/Gaia catalogs, rather than the ``main'' orbital period.

Additional checks for blended targets could use the expected amplitude of the variability. Even if the amplitude of the extra eclipses or pulsations is small in the TESS light curves, nearby, faint ($G\sim17$~mag) stars can still cause contamination if the amplitude of the variability in the blended variable is large. We do not compare the expected amplitudes here because of the challenges in comparing amplitudes from catalogs that use different filters.

Figure \ref{fig:param_hists} shows the distributions of the apparent $G$-band magnitudes and orbital periods. The majority of our targets are at periods shorter than a TESS sector, but we identify some extra-physics targets at longer periods using the ASAS-SN light curves. Many of these targets are bright enough for radial velocity followup, which could be used to confirm the triple/quadruple system candidates. 

We cross-match our catalog with Gaia Data Release 3 \citep[GDR3,][]{GaiaCollab2022} and the catalog of distances from \citet{BailerJones21}. We combine the color and distance information with extinction estimates from the {\tt mwdust} \citep{Bovy16} 3-dimensional `Combined19' dust map \citep{Drimmel03, Marshall06, Green19}. We use the MESA Isochrones \& Stellar Tracks \citep[MIST,][]{Choi16, Dotter16} to divide the color-magnitude diagram (CMD) into systems with main sequence and giant/subgiant primaries. Following \citetalias{Rowan22}, we only report the evolutionary state for systems with $\texttt{parallax\_over\_error} > 10$ and $A_V < 2.0$~mag. Figure \ref{fig:cmd} shows the extra-physics systems on a CMD colored by their variability type. 

Since rotational variability in particular can produce high-energy emission, we cross-matched our catalog with the HEASARC Master X-ray catalog\footnote{\url{https://heasarc.gsfc.nasa.gov/W3Browse/all/xray.html}} and the Swift-XRT Point Source Catalog \citep{Evans20}, both with a search radius of $10\farcs0$. In the HEASARC Master X-ray catalog, we find X-ray sources in ROSAT \citep{Voges99}, XMM-Newton \citep{Jansen01}, and Chandra \citep{Evans10} catalogs, which all have different localization errors, so X-ray sources at larger separations should be confirmed. In total, we find X-ray matches for \nXrayTotal{} targets, \nXraySpotted{} of which are spotted binaries. The remaining systems are in the \XrayGroups{} groups. Table \ref{tab:ep_table} reports the X-ray luminosity, $L_X$, computed using the \citet{BailerJones21} distance measurements, and the angular separation between the X-ray and optical position. For targets with multiple X-ray matches we use the flux corresponding to the longest exposure to compute $L_X$. We next discuss each group in order of their size.

\subsection{Spotted Stars} \label{sec:spotted_stars}

\begin{figure}
    \centering
    \includegraphics[width=\linewidth]{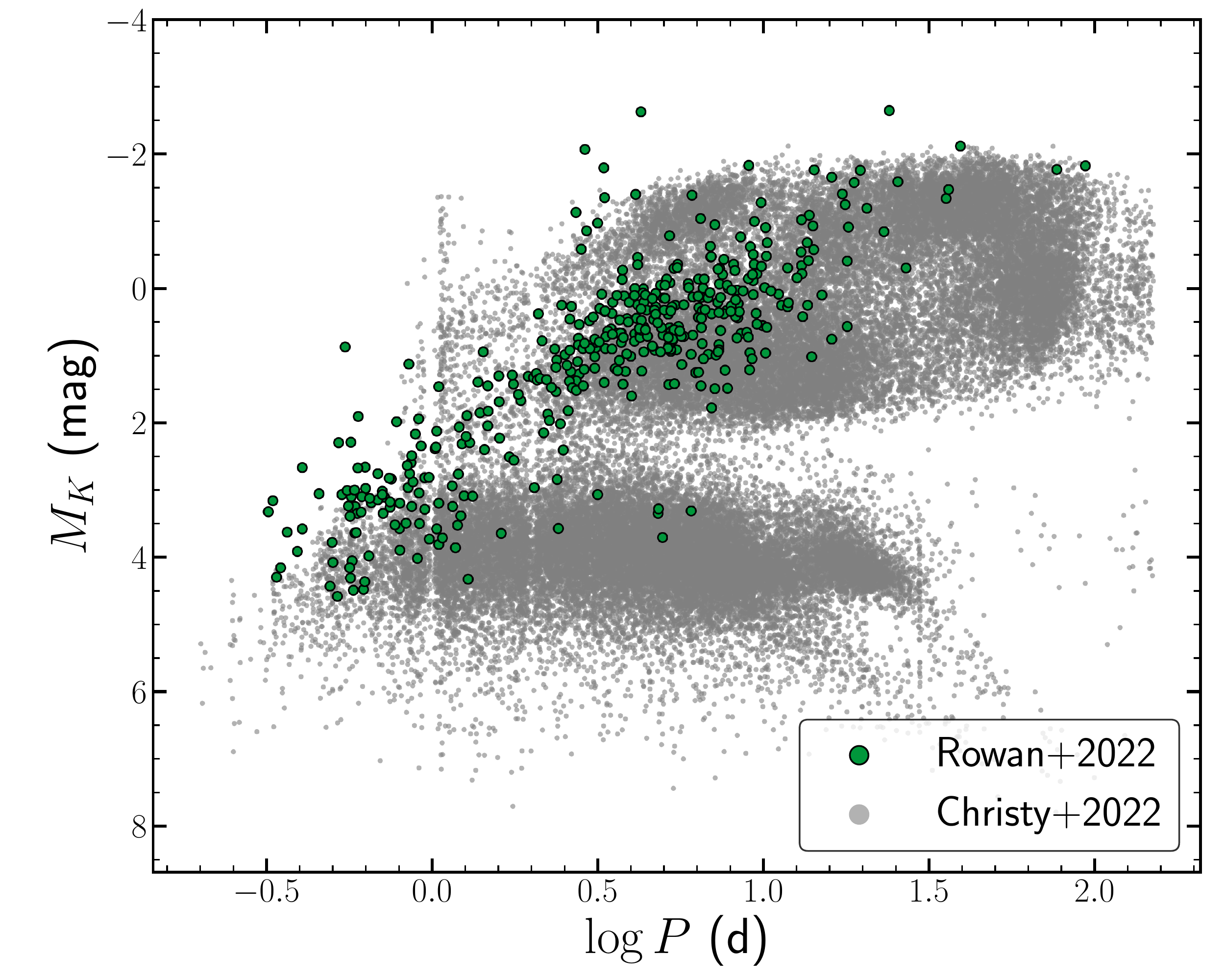}
    \caption{Distribution of $M_K$ and orbital period for the rotational variables from the ASAS-SN $g$-band variables catalog \citep{Christy22} and the \nSpotted{} targets in our catalog. Distinct clumps are found in the ASAS-SN $g$-band catalog, but the binaries in our catalog do not trace the clumps well.}
    \label{fig:spot_clumps}
\end{figure}

The distribution of the orbital periods of the spotted eclipsing binaries is bimodal (Figure \ref{fig:param_hists}). The short period sample is almost exclusively made up of systems with MS primaries, and the longer periods are SG and RG systems. The median orbital period of the MS spotted binaries is \medPeriodSpottedMS{}~days and the median period of the SG and RG systems is \medPeriodSpottedSGRG{}~days. This division by evolutionary state is relatively pure -- only \nSpottedMSlongP{} (\FracSpottedMSlongP{}\%) MS have periods $\log(P)>0.3$~days and \nSpottedSGRGshortP{} (\FracSpottedSGRGshortP{}\%) of SG and RG systems have periods shorter than $\log(P) < 0.3$~days. In total, \nXraySpotted{} of spotted binaries have X-ray detections within $10\farcs0$. Compared to the full sample of spotted binaries, the systems with X-ray emission tend to have shorter periods. The MS spotted binaries with X-ray emission have a median orbital period of \medPeriodSpottedXrayMS{}~days compared to \medPeriodSpottedMS{}~days and the median period of the SG/G spotted systems decreases from \medPeriodSpottedSGRG{}~days to \medPeriodSpottedXraySGRG{}~days.

Figure \ref{fig:spot_clumps} shows the distribution of extinction-corrected absolute $K$-band magnitude, $M_K$, and $\log P$ for the spotted binaries. \citet{Christy22} identified unique `clumps' of rotational variables in this parameter space. The physical properties and the differences between the clumps of rotational variables are explored in Phillips~et~al. ~(in~prep), but the spotted binaries in our catalog do not trace these clumps well.

We calculate rotation periods for the TESS light curves by manually masking the eclipses and running Lomb-Scargle periodograms  \citep{Lomb76, Scargle82} on the remaining light curves. In total, we compute rotation periods for \nSpotperiodtwo{} of the \nSpotted{} binaries, excluding systems where the orbital/rotational period is long compared to the length of a TESS sector or systematics in the TESS light curve limit determination of the period. We find that the spotted binaries are almost all near tidal synchronization, which is expected for the short period systems in our sample \citep[e.g.,][]{Lurie17}. Although we do observe some long-term modulations between TESS sectors (shown in Figure \ref{fig:appendix_spots}) that could be due to differential rotation \citep[e.g.,][]{Reinhold13}, the gaps between the TESS sectors limit our ability to determine these modulation periods. 

Three binaries have rotation to orbital period ratios $|\log({P_{\rm{rot}}/P_{\rm{orb}}})|>0.05$: Gaia DR3 5900431676207912448, 505860530871568512, and 1974523395151038208. Gaia DR3 5900431676207912448 is on the subgiant branch, and the rotation period of $P_{\rm{rot}}=0.82$~days is about 1/5 of the orbital period of $P_{\rm{orb}}=4.32$~days. The second target, Gaia DR3 505860530871568512, has {\tt parallax\_over\_error < 10}, so we have no estimate of the evolutionary state. The periods of $P_{\rm{rot}}=1.93$~days and $P_{\rm{orb}}=2.92$~days are in a near 2:3 ratio. Finally, Gaia DR3 1974523395151038208 is on the upper main sequence ($M_G\simeq-1.2$~mag, $G_{\rm{BP}}-G_{\rm{RP}}\simeq-0.18$~mag) and the periods are $P_{\rm{rot}}=3.89$~days and $P_{\rm{orb}}=3.15$~days. This system has ATLAS $\texttt{r1}=32\farcs4$, and nearby variable stars in the ATLAS (J328.6176+46.9429) and ZTF (J215427.72+465633.8) catalogs, but neither have periods matching the rotation period. 

Two targets, Gaia DR3 5201190414610275456 and 2191346572751253376, have $G_{\rm{BP}}-G_{\rm{RP}} > 1.5$~mag after correcting for extinction. These systems are likely sub-sub giants, which are redder and fainter than typical giants/subgiants. Sub-subgiants are expected to form through mass transfer, envelope stripping, and/or magnetic fields \citep{Mathieu03}. The orbital periods are 11.88 and 6.36 days, respectively, which are consistent with what is expected from a stripping interaction \citep{Leiner17}. Gaia DR3 5201190414610275456 is also identified as an X-ray source in Chandra \citep{Evans10}, XMM-Newton \citep{Webb20}, and Swift \citep{Evans20}. Both of these are promising targets for radial velocity followup to determine the masses and radii of sub-subgiants. 

\subsection{Extra Eclipses} \label{sec:extra_eclipses}

\begin{figure}
    \centering
    \includegraphics[width=\linewidth]{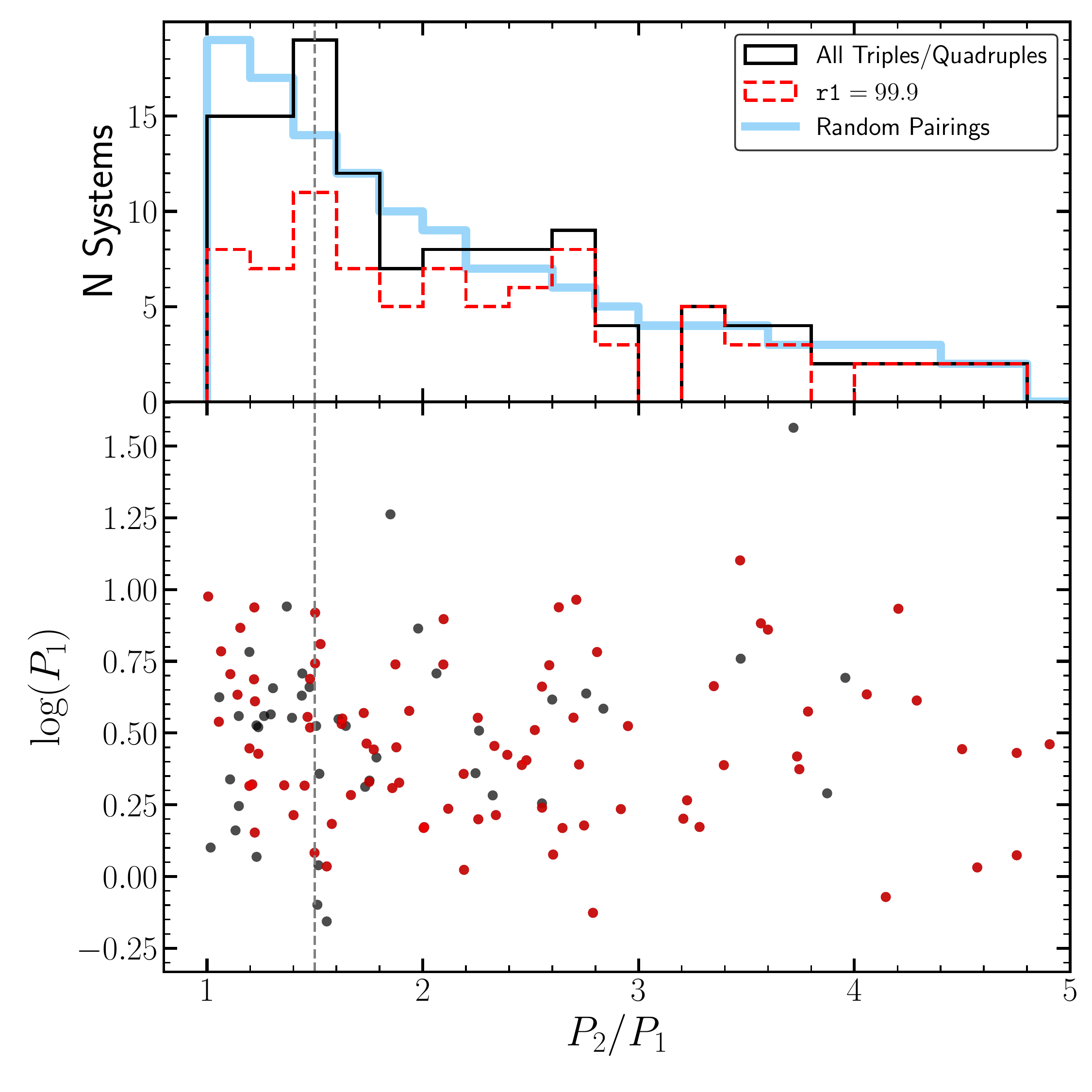}
    \caption{Orbital periods $P_1$ and $P_2$ for triple/quadruple systems. For the purposes of this figure, $P_2$ is defined as the longer of the two orbital periods. Top: distribution of $P_2$/$P_1$. Systems with $\texttt{r1}=99$ are shown in red. Bottom: Distribution of $P_2$/$P_1$ and $\log(P_1)$, showing there are some systems with period ratios exactly at 3:2.}
    \label{fig:p2_result}
\end{figure}

Unlike the spotted stars, which have a bimodal distribution of $\log P$, the distribution of $\log P$ for the extra-eclipsing systems is unimodal with a median period of \medPerEE{}~days. Figure \ref{fig:skymap} shows that these systems are also more concentrated near the Galactic plane near the spotted stars, similar to the quadruple systems in \citet{Zasche22}. The extra-eclipsing systems are distributed across the main sequence, although they are typically of earlier types than the spotted stars. There are \nEERGSG{} systems with extra eclipses on the subgiant and red giant branches.

To determine the orbital period of the tertiary or second binary, $P_2$, we start by folding the ASAS-SN $g$-band and TESS light curves at the original orbital period, $P_1$. We then use the \PHOEBE{} geometry estimator, which combines a two Gaussian model with a cosine term to estimate the orbital eccentricity, argument of periastron, and time of superior conjunction, to construct an analytic model. We try four different methods of subtraction/masking before searching for the second period:

\begin{enumerate}
    \item Use the $g$-band analytic model to define the times of primary and secondary eclipse and mask out these times in the TESS light curve;
    \item Use the TESS analytic model to define and mask out times of primary and secondary eclipse;
    \item Subtract the analytic model from the TESS data; or
    \item Subtract the analytic model from the TESS data and mask out the eclipse
\end{enumerate}

We then use the {\tt astropy} implementation of the Box Least Squares (BLS) periodogram \citep{Kovacs02, Astropy13} to determine $P_2$. We phase-fold the light curve at $P_2$ and check that the period is reasonable and double the period when necessary. For some quadruple systems where the second binary is in contact, it is more effective to run a Generalized Least Squares (GLS) periodogram \citep{Lomb76, Scargle82}. While these four methods often produce the same $P_2$, there are some cases where one method is more effective. For instance, for light curves with non-flat out of eclipse shapes, methods (iii) and (iv) tend to be more effective because the periodogram power at the main binary orbital period is reduced. During this visual inspection process we also identify targets that require sigma-clipping due to TESS systematics before the period search is performed.

For systems where $P_2$ is greater than the length of a TESS sector we are unable to calculate $P_2$ if only one sector of data is available. There are also some systems where the systematics in the TESS light curve limit our ability to identify $P_2$ even after clipping bad data points. Finally, there are a small number of systems where there are additional effects in the light curve, such as reflection, that are not represented by the analytic model. In total, we report values of $P_2$ for \nEEperiodtwo{} systems in Table \ref{tab:ep_table}. 

Figure \ref{fig:p2_result} shows the distribution of period ratios where $P_2$ is defined to be the longer of the two periods. For comparison, we randomly pair the $P$ and $P_2$ orbital periods and show their distribution in period ratio as the blue line in Figure \ref{fig:p2_result}. There is a clear overabundance in systems near a period ratio of 3:2 in the sample, even when only including systems with $\texttt{r1}=99.9$. \citet{Zasche19} also identified an excess of systems at a 3:2 period ratio. The bottom panel of Figure \ref{fig:p2_result} shows that these systems fall almost exactly at $P_2/P_1=1.5$, which is expected for orbital resonance. 

However, we note that two of the binaries are apparent blends with each other. Gaia DR3 5618940344244565504 and Gaia DR3 5618940653482188160 are separated by $\sim60\arcsec$, and both TESS light curves have eclipses with periods $P_1=5.529231$~d and $P_2=8.300885$~d, so $P_2/P_1=1.5013$. 

We cross-match the systems with extra eclipses with catalogs of triples and doubly-eclipsing binaries from \citet{Zasche19}, \citet{Zasche22}, \citet{Borkovits22}, \citet{Kostov22}, \citet{Rappaport22} and Table 1 of \citet{Borkovits20}. In total, \nEEknown{} (\FracEEknown{}\%) of our extra-eclipsing systems have been included in at least one of these catalogs\footnoteEEknown{}. Table \ref{tab:ep_table} includes the citations for each binary. Out of these \nEEknown{} systems, \EEKnownCont{} have nearby variable stars listed in Table \ref{tab:contamination_table}. One of these is CzeV343, which was discovered by \citet{Cagas12} and studied in detail in \citet{Pejcha22} and is clearly not a blend. The nearby variable star, ATLAS J087.1044+30.9502, is separated by $13\farcs7$ and is listed as a dubious variable. The triple system TIC 52041148, is also returned in our cross-match to variables catalogs, and is unlikely to be a blend since it shows evidence of eclipse timing variations and eclipses of the teritary in the ASAS-SN photometry \citep{Borkovits22}. The remaining systems in the known triple/quadruple catalogs that have variable matches in Table \ref{tab:contamination_table} are in the \citet{Kostov22} and \citet{Zasche22} catalogs. Only one of these, TIC 251757935, has a variable star match to $P_2$ in the ATLAS catalog, but the separation between the Gaia coordinates and the ATLAS variable is $<0\farcs1$, suggesting that ATLAS has recovered the second orbital period and this is a true triple/quadruple system. 

\subsection{Reflection Effects} \label{sec:reflection_systems}

Mutual surface heating in close binary systems can alter the out-of-eclipse shape of the light curve. This reflection effect is generally more prominent for systems with shorter periods and large temperature differences. In our catalog, the median period of all \nEP{} systems is \medPeriod{}~days, but the \nREFL{} reflection systems have a median period of \medPerREFL{}~days. Of the \nREFLQuality{} reflection systems with Gaia $\texttt{parallax\_over\_error} > 10$ and $A_V<2.0$~mag, \nREFLms{} have MS primaries and \nREFLsgrg{} have SG/RG primaries.


Three systems, Gaia DR3 6144569024718252544, 1375814952762454272, and 6652952415078798208 are much bluer than the binary main sequence with  with $G_{\rm{BP}}-G_{\rm{RP}}\simeq-0.4$~mag (see Figure \ref{fig:cmd}). The first two have $M_G\simeq4.1$~mag and $P<0.2$~days and are likely subdwarf B and M dwarf (sdB+dM) systems \citep[e.g.,][]{Dai22}. Gaia DR3 6652952415078798208 is more luminous, with $M_G=2.3$~mag, and has a longer period of $P=0.36$~days, but it is likely also an sdB+dM binary, rather than an sdO+dM binary based on the color classification schemes given in Table 1 of \citet{Geier20}. 

This system also has additional scatter in the light curve that could be due to pulsations. We used the \PHOEBE{} geometry estimator to mask the eclipses in the TESS light curve and then fit a sinusoid to subtract off the reflection effect signal. We use a Lomb-Scargle periodogram and find additional variability at $P=0.575455$~days with false-alarm probability $<10^{-7}$. This frequency is lower than typically observed for pulsating sdB stars \citep{Reed14, Baran19}, so the nature of this variability is unclear.

Hot subdwarfs are likely to form through binary evolution with phases of common envelope evolution and Roche lobe overflow \citep[e.g.,][]{Han02}, although single-star evolutionary channels \citep[e.g.,][]{Dcruz96} and hierarchical triple models \citep[e.g.,][]{Preece22} have been proposed. Radial velocity followup of these targets can be used to characterize the masses and radii of both components and evaluate formation models. 

\subsection{Regular Pulsators} \label{sec:regular_pulsators}

\begin{figure}
    \centering
    \includegraphics[width=\linewidth]{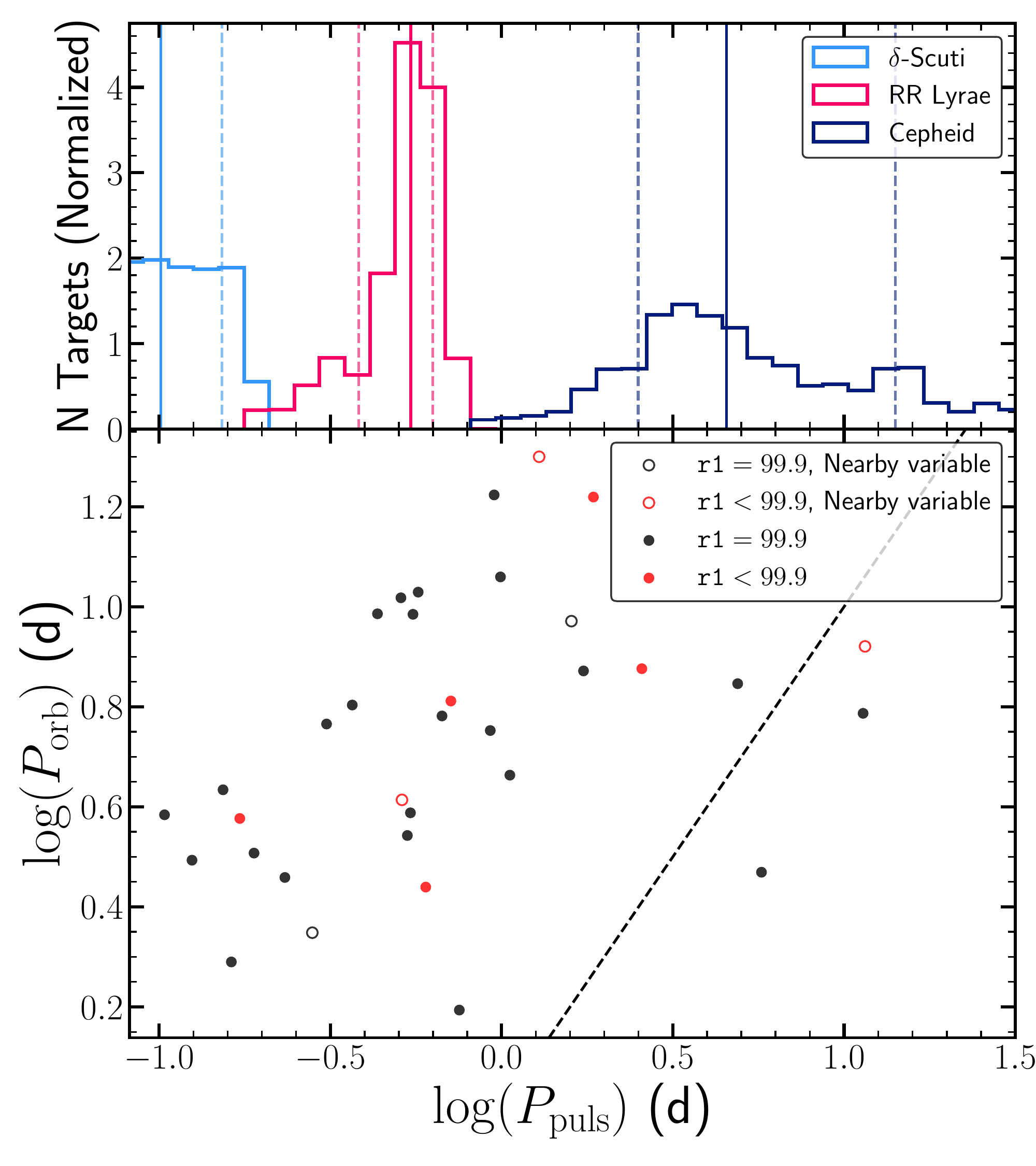}
    \caption{Pulsation periods for the \nRP{} eclipsing systems with periodic pulsations. The top panel shows the normalized distribution of $\delta$-Scuti, RR Lyare, and Cepheid variable periods from ASAS-SN with classification probabilities $>0.9$ from \citet{Jayasinghe21}. The vertical solid lines show the median $\log P$ of the distributions and the dashed lines show the 16th and 84th quantiles of the distributions. Systems above the black line have pulsation periods longer than the orbital period. Systems with $\texttt{r}1<36\arcsec$ are shown in red, and systems with nearby variable stars in Table \ref{tab:contamination_table} are shown as open circles.}
    \label{fig:pulsation_periods}
\end{figure}

\nRP{} systems have additional periodic variability in their light curves that is due to pulsations rather than extra eclipses. We use the procedures described in Section \ref{sec:extra_eclipses}, substituting the Generalized Lomb Scargle periodogram \citep{Lomb76, Scargle82} rather than a BLS, to calculate the period. We are able to calculate the pulsation period for all but 2 systems, and report the
pulsation periods in Table \ref{tab:ep_table}. Figure \ref{fig:pulsation_periods} shows the distribution of pulsation and orbital periods compared to the distribution of pulsational variables in \citet{Jayasinghe21}. 


The majority of targets have $\delta$-Scuti, RR Lyrae, or Cepheid-like pulsations. We identify four systems with Cepheid pulsations: Gaia DR3 6140491142248459776 ($P_{\rm{orb}}=2.95$~d), 2070234748476679424 ($P_{\rm{orb}}=9.36$~d), 2952256516644702720 ($P_{\rm{orb}}=7.02$~d), and 5350308140079159168 ($P_{\rm{orb}}=16.58$~d). The smallest Cepheids are expected to have radii $\sim 30$~R$_\odot$ \citep{Gieren99} and masses 3--5~M$_\odot$. Even if the companion was a low-mass main sequence star, which is unlikely given the relative eclipse depths, the orbital period based on Kepler's third law must be $\gtrsim 10$~days. Since these systems are detached with minimal ellipsoidal effects in the light curves, the periods would have to be even larger, suggesting these targets are blends. However, only one of these targets, Gaia DR3 5350308140079159168, has $\texttt{r}1=29\farcs8$. 

Of these four systems, only Gaia DR3 2070234748476679424 with $P_{\rm{puls}}=1.6$~days has nearby variable stars in the ZTF and ATLAS catalogs. ZTF J204433.65+452440.5 is separated by $16\farcs7$, has a $g$-band period of $0.87$~days, and is classified as an RS CVn \citep{Chen20}. We downloaded and phase-folded the ZTF light curve at our derived $P_{\rm{puls}}$ and do not find pulsational variability consistent with the TESS period. The ATLAS variables catalog \citep{Heinze18} includes J311.1366+45.4141, which is separated by $17\farcs2$. The Lomb-Scargle period is $0.39$~days, and the system is listed as a ``dubious'' variable. This system must be a blend given the orbital period and the typical radii of Cepheids, but nearby matches to variable catalogs do not contain systems with the expected period.

Gaia DR3 6727189722220410752 and 3028351483024584832 are apparent RR Lyrae eclipsing binaries with ATLAS $\texttt{r1}=99.9$. Neither of these targets have nearby variable stars $<21\arcsec$ in the Gaia, ATLAS, ZTF, or ASAS-SN variable star catalogs. We classify the later system as a first overtone RR Lyrae (RRC) because of the short pulsation period $P_{\rm{puls}} = 0.31$~days and the nearly sinusoidal nature of the light curve. Neither of the ASAS-SN light curves of these targets shows similar pulsational variability at these periods. 

We expanded the search radius from $21\arcsec$ to $60\arcsec$ and repeated the cross match for both targets. For Gaia DR3 6727189722220410752, we find an RR Lyrae separated by $57\farcs1$ in the ASAS-SN $V$-band catalog \citep{Jayasinghe21} with the same pulsation period $P_{\rm{puls}} = 0.57$~days that we recovered from the TESS light curve. We find no nearby contaminating variable stars within $1\arcmin$ for 3028351483024584832, so we further extend the search radius to $3\arcmin$. We find no matches in the ASAS-SN variable catalog, but identify ZTF~J073112.04-152905.2, separated by $2\farcm8$. \citet{Chen20} identified this system as a contact binary with a period of $0.32$~days, which is consistent with our derived pulsation period. The amplitude of the variability in the TESS light curve is small ($\sim 0.025$~mag). However, ZTF~J073112.04-152905.2 is $g\sim17.59$~mag and \citet{Chen20} reports and amplitude $\Delta g=0.452$~mag. 

Whether this system is best classified as a contact binary or a pulsator, it is likely that the extra variability in the TESS light curve of 3028351483024584832 is due to a blended variable. Both of these variables demonstrate the wide separations where blending from variable stars with high amplitudes can contaminate TESS light curves.

We find two targets that we label as $\beta$-Cephei pulsations. Gaia DR3 3123204972458091008 is in a $P_{\rm{orb}}=6.05$~day binary with a pulsation period of $P_{\rm{puls}}=0.67$~days. Gaia DR3 5310247781838683648 is in a $P_{\rm{orb}}=1.56$~day binary and has a pulsation period of $P_{\rm{puls}}=0.75$~days. Both targets are found on the upper main sequence and are consistent with A or B type primaries. The pulsations of Gaia DR3 3123204972458091008 are highly regular both in their arrival time and amplitude, and are similar to those of $\eta$~Ori \citep[][Figure 1]{Southworth22} although only one Sector of TESS data is available. The binary itself is also eccentric, $e>0.2$ based on the \PHOEBE{} geometry estimator fit. On the other hand, the Gaia DR3 5310247781838683648 pulsations change in amplitude across the single Sector of available TESS data and are more similar to 16 Lac \citep{Southworth22}. Both of these binaries are promising targets for astereoseismic modeling with additional constraints from spectroscopic orbit results. Such systems can be used to validate and improve interior structure models for massive stars and the impact of close binary companions \citep[e.g.][]{Soydugan06a, Schmid16, Tkachenko20}.

\subsection{Stochastic Variability} \label{sec:stochasic_variability}

We identify \nSto{} targets with stochastic variability in their light curves. These were identified and distinguished from the regular pulsators and $\gamma$-Dor pulsators by inspecting the unfolded TESS light curves. The variability has typical timescales of hours to days, and is consistent with the stochastic low frequency variability that arises from from turbulent core convection driven gravity waves in massive stars \citep{Bowman19, Bowman20, Southworth22}. Figure \ref{fig:appendix_sto_var} shows examples of their TESS light curves. 

We only report the evolutionary state for \nStoVarState{} of these targets. The rest have extinction $A_V > 2.0$ (\nStoVarHighAV{} systems), Gaia $\texttt{parallax\_over\_error} < 10$ (\nStoVarBadPLX{} systems) or both (\nStoVarBoth{} system). Of the systems with reported evolutionary states, 7 have absolute magnitudes consistent with OBA MS stars. One system, Gaia DR3 5560556684818159872, is on the giant branch, which might indicate a different origin of variability compared to the rest of the targets. 

\subsection{\texorpdfstring{$\gamma$-Dor Pulsators}{Gamma-Dor Pulsators}}  \label{sec:gamma_dor_pulsators}

\begin{figure}
    \centering
    \includegraphics[width=\linewidth]{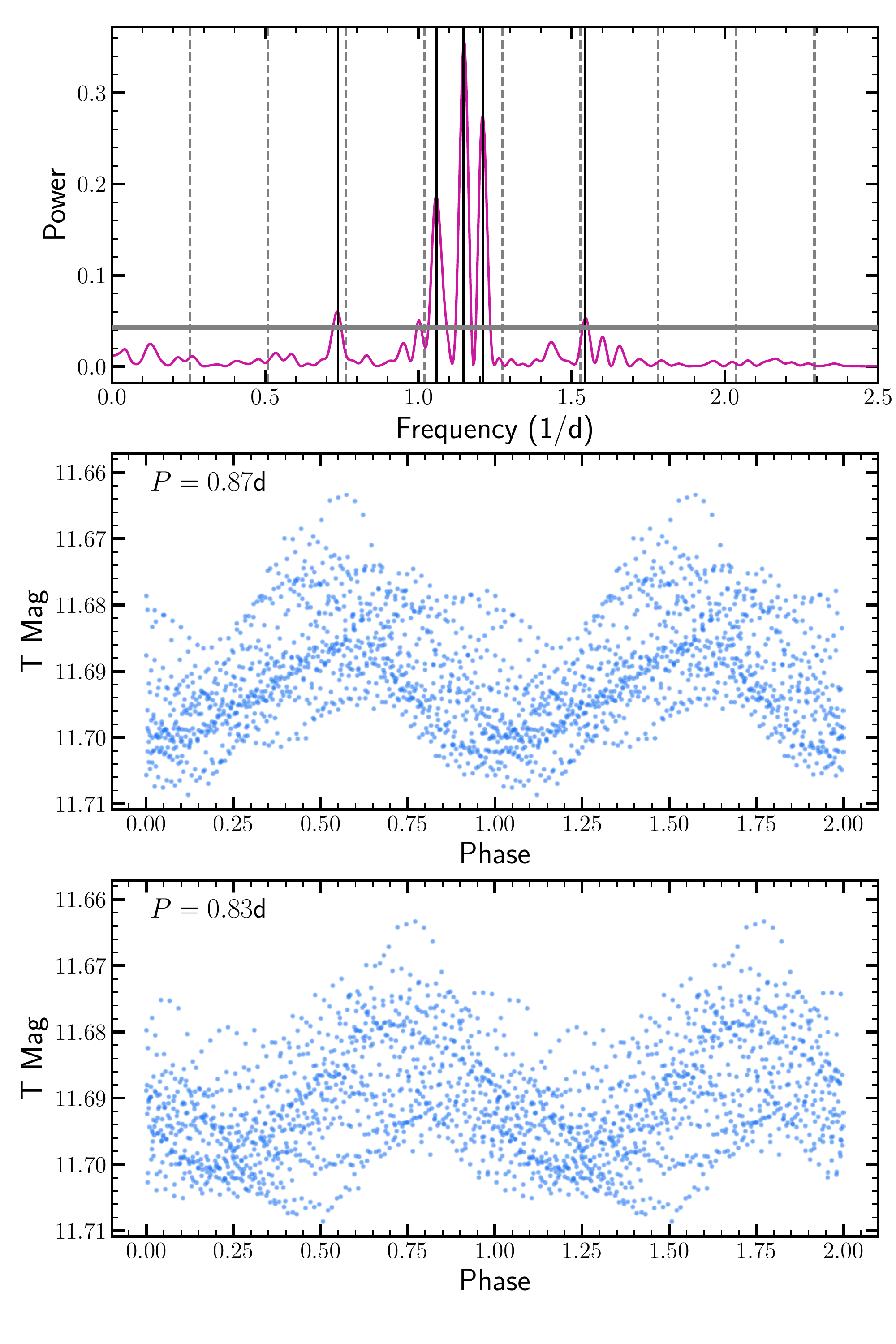}
    \caption{Example of a frequency search for the $\gamma$-Dor eclipsing candidate Gaia DR3 263700752773032448. Top: the Lomb-Scargle periodogram after masking the eclipses. The vertical dashed gray lines show harmonics of the orbital frequency, and the solid lines show the five frequencies reported in Table \ref{tab:gammador_table}. The horizontal gray line shows the false alarm level of $10^{-5}$. Middle: the TESS light curve folded at the frequency with the highest power. Bottom: the TESS light curve folded at the frequency with the second highest power.}
    \label{fig:gammador_example}
\end{figure}

\begin{table}
    \centering
    \caption{Pulsation frequencies for the \nIRP{} eclipsing $\gamma$-Dor variable candidates.}
    \sisetup{table-auto-round,
         group-digits=false}
    \setlength{\tabcolsep}{3pt}
    \begin{center}
        \begin{tabular}{l S[table-format=2.3] S[table-format=1.3] S[table-format=1.3] S[table-format=1.3] S[table-format=1.3] S[table-format=1.3]}
\toprule
         {Gaia DR3} &  {$\nu_{\rm{orb}}$} &  {$\nu_1$} &  {$\nu_2$} & {$\nu_3$} & {$\nu_4$} & {$\nu_5$} \\
{} & {(1/d)} & {(1/d)} & {(1/d)} & {(1/d)} & {(1/d)} & {(1/d)}\\ 
\midrule
5808714298849592192 &            0.518587 &   1.266034 &   1.130371 &  0.227574 &   2.30594 &   2.16883 \\
 443999635867486720 &            0.518191 &   1.179542 &   1.052688 &  1.443481 &  1.310489 &           \\
5331653898816403584 &            0.441592 &   0.880260 &   0.718965 &  0.953472 &  0.561101 &  0.598597 \\
 248169674294594816 &            0.381381 &   0.913561 &   0.977446 &  0.815603 &  0.764495 &  1.058367 \\
 434846854403757184 &            0.289085 &   0.857623 &   0.702932 &  0.798476 &  0.643786 &  0.284358 \\
 263700752773032448 &            0.254719 &   1.146765 &   1.211032 &  1.058398 &  0.737063 &  1.544417 \\
1999637031003866368 &            0.244145 &   1.478099 &   1.710215 &  1.543792 &  1.648901 &  1.815324 \\
5693600963342153088 &            0.220179 &   0.439912 &   0.871737 &  0.806902 &   0.38364 &  0.609271 \\
 351850661555541504 &            0.191335 &   0.647220 &   0.587803 &  1.084359 &  0.706637 &  0.528387 \\
5248678051242539136 &            0.188473 &   0.712255 &   0.491157 &  0.545574 &  0.459782 &  0.637248 \\
3103470323585557376 &            0.184937 &   0.608810 &   0.787458 &  0.674628 &  0.970806 &           \\
  23092256739885312 &            0.175998 &   0.600004 &   0.662955 &  0.497708 &  0.552791 &  1.166566 \\
4532112201568025728 &            0.168784 &   0.554279 &   0.507039 &  0.986297 &  0.933916 &  0.432384 \\
3143993576241783040 &            0.158851 &   1.546512 &   1.214204 &  0.643315 &  0.762605 &   0.23432 \\
3249778590678600448 &            0.126088 &   0.376776 &   0.483626 &   0.43512 &  0.627504 &  0.317339 \\
5288047130219214080 &            0.122173 &   0.526709 &   0.365001 &  0.464402 &  0.417744 &  0.386253 \\
 435896716209335040 &            0.120582 &   1.039225 &   1.234954 &           &           &           \\
\bottomrule
\end{tabular}

    \end{center}
    \label{tab:gammador_table}
\end{table}

We identify \nIRP{} systems with $\gamma$-Dor pulsations, all of which are near $G_{\rm{BP}}-G_{\rm{RP}}\sim0.5$~mag and $M_G\sim2.5$~mag on the CMD (Figure \ref{fig:cmd}) as expected from catalogs of $\gamma$-Dor pulsators \citep[e.g.,][]{Li20}. All of the $\gamma$-Dor systems except for Gaia DR3 5331653898816403584 have ATLAS $\texttt{r1}=99.9$ and all have orbital periods $P<10$~days (Figure \ref{fig:param_hists}). Since many of these systems have multiple frequency components and the amplitudes of the pulsations change over the course of a single binary orbit, we separated them from the regular pulsators. We perform a similar period search as for the regular pulsators described in Section \ref{sec:regular_pulsators}, but identify multiple frequencies for most of our targets above a false alarm level of $10^{-5}$. We also only considered pulsation periods shorter than 5~days based on typical periods for $\gamma$-Dor pulsators \citep[e.g.,][]{Henry11}.

Figure \ref{fig:gammador_example} shows an example periodogram and light curves for Gaia DR3 263700752773032448. The top panel shows the Lomb-Scargle periodogram where the highest-power frequencies are not consistent with harmonics of the orbital frequency (gray dashed lines). The bottom two panels show the TESS light curve folded at the two best frequencies. In total, we detect six significant pulsation frequencies for this target. Table \ref{tab:gammador_table} reports the five best pulsation frequencies for all targets. Since $\gamma$-Dor pulsations are gravity-mode, or g-mode, pulsations \citep{Kaye99}, they can be used to study the interior structures of stars \citep{VanReeth16} with additional constraints from the masses and radii based on the eclipses \citep[e.g.,][]{Southworth22_gammador}.

\subsection{Other Targets} \label{sec:other_targets}

\begin{figure*}
    \centering
    \includegraphics[width=\linewidth]{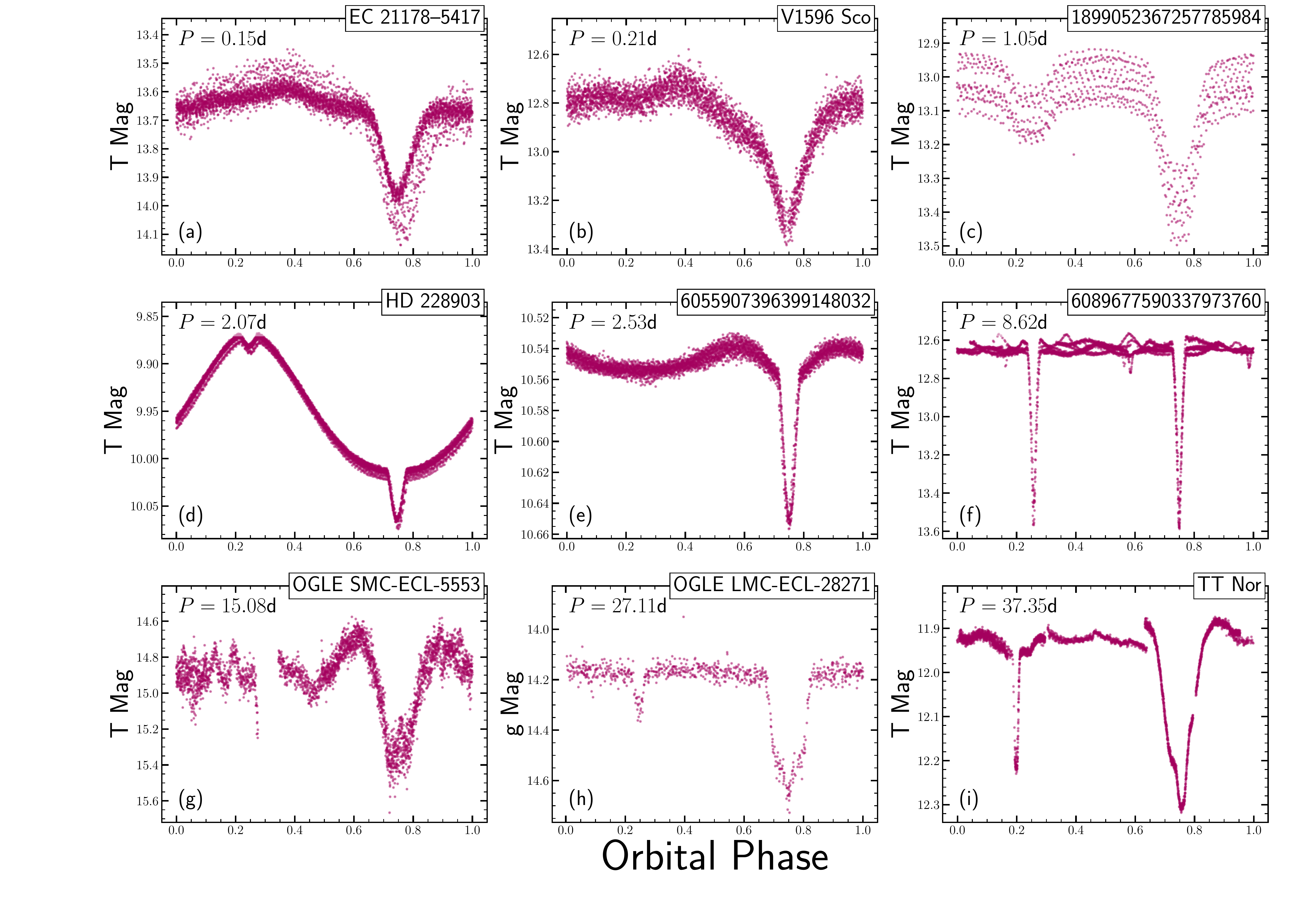}
    \caption{Same as Figure \ref{fig:panel_plot}, but for systems that do not fit into one of the main classifications. The properties of these systems are described in Section \ref{sec:other_targets}. The Gaia DR3 source or other identifier are reported in the top right of each panel.}
    \label{fig:other_panel}
\end{figure*}

\begin{figure}
    \centering
    \includegraphics[width=\linewidth]{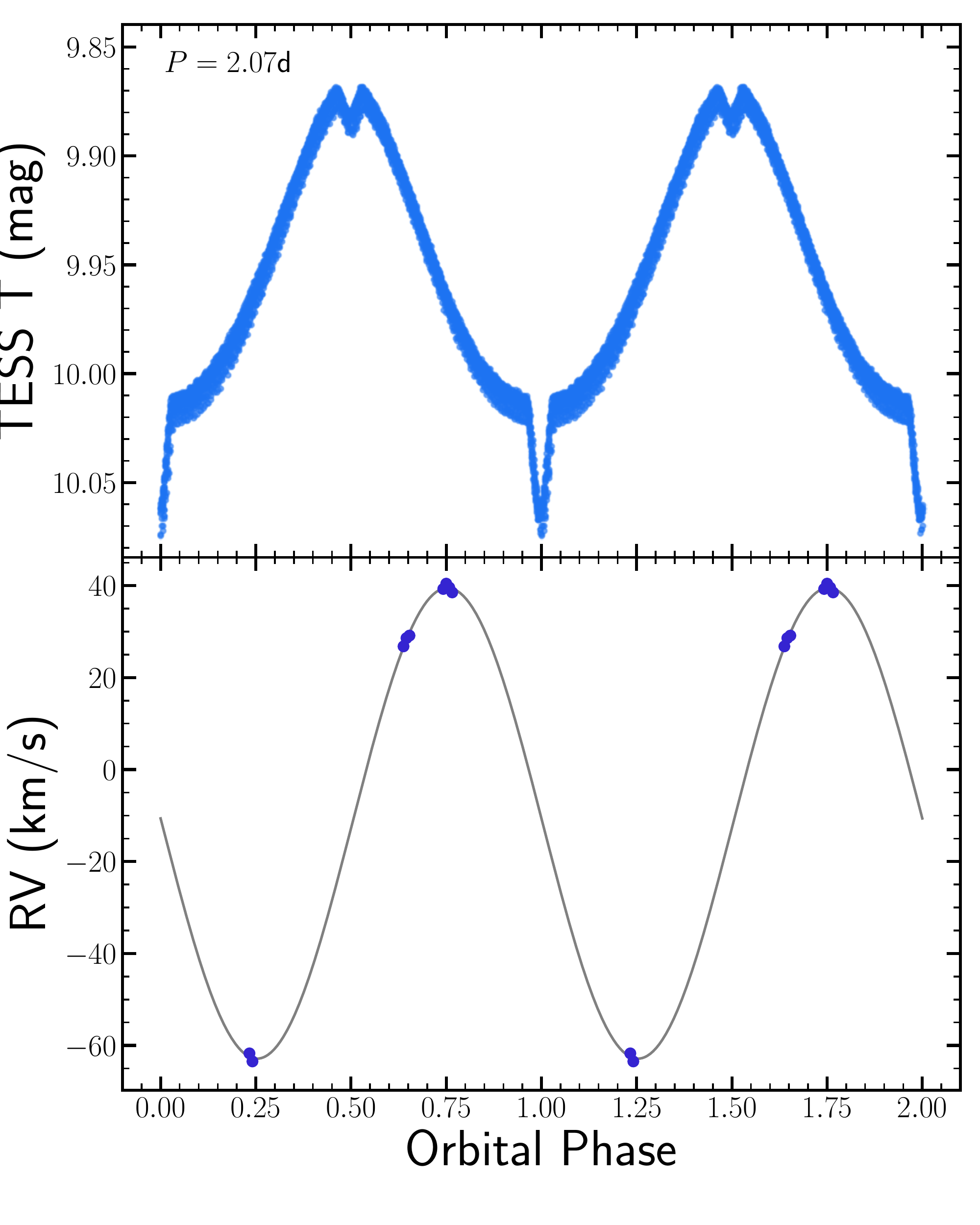}
    \caption{Light curve and LAMOST radial velocity curve of HD 228903, an eclipsing subdwarf system, phased such that the deeper eclipse is at phase $0.0$. The gray line shows a circular orbit fit to the RV data.}
    \label{fig:eclipsing_subdwarf_rv}
\end{figure}

\begin{figure*}
    \centering
    \includegraphics[width=0.75\linewidth]{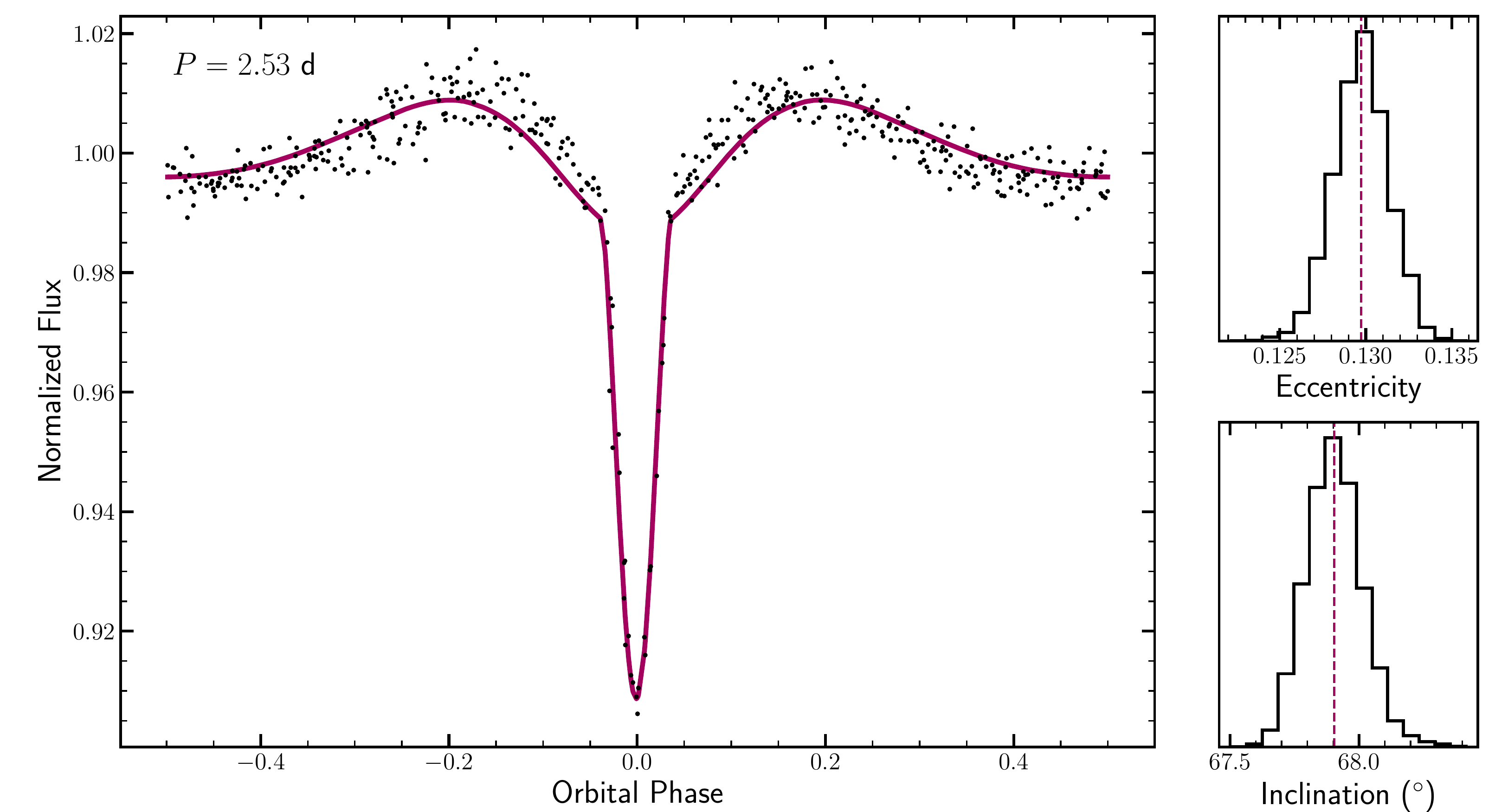}
    \caption{TESS light curve of Gaia DR3 6055907396399148032 and the MCMC posteriors for the eccentricity and inclination. The geometry of this system prevents detection of a secondary eclipse, so the system geometry is determined from the distortions near the primary eclipse.}
    \label{fig:53788679_lc}
\end{figure*}

There are \nOther{} systems that do not fit into these categories and are shown in Figure \ref{fig:other_panel}. Here we briefly describe their properties and offer possible classifications. 

\subsubsection{White Dwarf+MS Disk Systems} \label{sec:white_dwarf_disk_systems}

Gaia DR3 6463668670854493952 (EC~21178--5417, Figure \ref{fig:other_panel}a) and 6031156290142441344 (V1596~Sco, Figure \ref{fig:other_panel}b) are found between the white dwarf cooling sequence and the main sequence on the CMD (Figure \ref{fig:cmd}). Both have orbital short orbital periods, $P=0.15$ and $P=0.21$~days, respectively. The light curves show a single eclipse deep eclipse with asymmetric ingress and egress shapes.

EC~21178--5417 is found in a number of CV catalogs \citep{Stobie97,Downes01, Ritter03, Ak08} and is typically classified as a nova-like variable. Photometric observations of this system have identified quasi-periodic oscillations (QPOs) and dwarf-nova oscillations \citep[DNOs,][]{Warner03} in addition to the eclipses. Spectroscopic observations identified double-peaked HeII emission lines, suggesting an inclined accretion disk \citep{Khangale20}.

We did not find V1596~Sco in any CV catalogs, but the system is likely also a nova-like variable based on the similarities to EC~21178--5417. V1596~Sco was included in the \citet{Shi22} catalog of pulsating eclipsing sources in TESS. We mask the eclipses in the TESS data and recover variability at the same period of $P=0.0513$~days. This period is about a factor of 10 longer than the QPOs and more than 100 times longer than the DNOs identified in EC~21178--5417 by \citet{Warner03}, and further observations are needed to identify the origin of this variability. 

\subsubsection{Ellipsoidal triple} \label{sec:ellipsoidal_triple}

Gaia DR3 1899052367257785984 (Figure \ref{fig:other_panel}c) is in a $1.05353$~day orbit detected in ASAS-SN \citep{Jayasinghe21}, ZTF \citep{Chen20}, and ATLAS \citep{Heinze18}. The TESS light curve shows additional variability at $9.09$~days that is consistent with ellipsoidal modulations. The amplitude of this variability is about half of the amplitude of the primary eclipse, but no variability is seen in the ASAS-SN, ZTF, or ATLAS light curves. ATLAS reports an $\texttt{r1}=16\farcs8$, suggesting this target is likely a blend. 

\subsubsection{Eclipsing Subdwarf} \label{sec:eclipsing_subdwarf}

Gaia DR3 2068355163403333504 (HD 228903) has a light curve consistent with eclipsing subdwarfs \citep[e.g.,][]{Ostensen10} and shows a strong reflection effect in addition to shallow, short duration eclipses evenly spaced in orbital phase (Figure \ref{fig:other_panel}d). The orbital period, $P=2.07231$~days is about $\sim1/4$ of the period reported in the ASAS-SN variable stars catalog \citep{Jayasinghe21}, and this target is not included in the ZTF or ATLAS variability catalogs, despite being in the northern hemisphere. This system was identified in the WISE catalog of variable stars with a period of 4.1443471~days in the miscellaneous category \citep{Chen18}. The WISE W1 and W2 light curves both show the reflection effect, but the eclipses are not visible. On the CMD, this target appears near the upper main sequence ($M_G\simeq-0.4$~mag, $G_{\rm{BP}}-G_{\rm{RP}}\simeq0.3$~mag) but it is likely evolving along the horizontal branch given the large temperature difference and similarities to subdwarf binaries and HW Virginis systems. HD 228903 has been detected as an X-ray source by XMM-Newton \citep{Traulsen20} and as UV source by Swift \citep{Yershov14}. 
  
HD 228903 was identified as a spectroscopic binary in the Large sky Area Multi-Object fiber Spectroscopic Telescope \citep[LAMOST,][]{Cui12} medium resolution survey \citep[MRS,][]{Zhang22} and nine radial velocity measurements are available. We use the photometric period and ephemeris to phase the radial velocity data as shown in Figure \ref{fig:eclipsing_subdwarf_rv}. Since the light curve suggests a circular orbit, we fixed the eccentricity to zero and found a velocity semi-amplitude of $K=51.2\pm0.4$~km/s and a mass function of $f(M)=0.029$~M$_\odot$. As expected, the maximum radial velocity occurs at phase 0.75.

\subsubsection{Eccentric inclined system} \label{sec:eccentric_inclined}


Gaia DR3 6055907396399148032 is a main sequence binary with an orbital period of $P=2.53314$~days. The light curve (Figure \ref{fig:other_panel}e) shows a single deep eclipse with a parabolic out of eclipse shape. Since the light curve maxima occur immediately before and after the eclipse, this system is probably in a inclined eccentric orbit without a secondary eclipse. We use {\tt emcee} \citep{ForemanMackey13} and \PHOEBE{} \citep{Prsa05, Prsa16, Conroy20} to fit the TESS light curve. To reduce computation time, we use only every tenth point in the TESS light curve. We start by following the optimization procedure described in \citetalias{Rowan22}, and then run the MCMC chain for 50000 iterations with a 3000 iteration burn in period. Figure \ref{fig:53788679_lc} shows the light curve sampled from the MCMC posteriors and the distributions of eccentricity and inclination. The estimated eccentricity and inclination are $e=0.130\pm0.002$ and $i=67.9\pm0.1^{\circ}$. The tidal distortion near periastron provides the constraints on the inclination and eccentricity. However, similar systems observed with lower photometric precision may be incorrectly assumed to be at twice the period and in a circular orbit. Radial velocity observations of this system can confirm the orbital period. 

\subsubsection{Extra-Eclipsing $\gamma$-Dor system} \label{sec:ee_gammador}


Gaia DR3 6089677590337973760 shows additional eclipses and $\gamma$-Dor-like pulsations (Figure \ref{fig:other_panel}f). We start by masking the deeper eclipses and identify a periodicity near $\sim24.2$~days corresponding to the additional pair of eclipses. Since data is only available for two TESS sectors, additional observations are needed to validate and further constrain this period. We then mask out the second pair of eclipses and use a Lomb-Scargle periodogram to identify the pulsation frequencies. The three dominant frequencies are at $\nu_1 = 0.578$, $\nu_2=0.430$, and $\nu_3=0.481$~d$^{-1}$. ATLAS reports $\texttt{r1}=4\farcs9$, but this target has declination $\textrm{DEC}=-51.1^{\circ}$, so it is outside the footprint of the ATLAS and ZTF variable catalogs. However, the $G=12.8$~mag variable Gaia DR3 6089677586043502848 is classified as an eclipsing binary with a period of $P=1.5256$~days. If Gaia DR3 6089677586043502848 is the blended target producing additional eclipses, the incorrect period in the Gaia catalog may come from the few epochs of observation available in the Gaia photometry.

To determine which eclipsing binary is more likely to have the $\gamma$-Dor pulsating component, we can consider their CMD positions. Gaia DR3 6089677590337973760 has absolute magnitude and color $M_G\simeq3.6$~mag, $G_{\rm{BP}}-G_{\rm{RP}}\simeq0.65$~mag. This is slightly fainter and redder than the majority of $\gamma$-Dor pulsators \citep[e.g. Figure 3 of][]{Sepulveda22}. The blended eclipsing binary, Gaia DR3 6089677586043502848 has absolute magnitude and color $M_G\simeq1.6$~mag, $G_{\rm{BP}}-G_{\rm{RP}}\simeq0.35$~mag, which is brighter than the $\gamma$-Dor pulsators in our catalog (Figure \ref{fig:cmd}), but is consistent with the brightest $\gamma$-Dor pulsators in the \citet{Li20} catalog. The nearby binary Gaia DR3 6089677586043502848 is therefore more likely to host the $\gamma$-Dor pulsator.

\subsubsection{Eclipsing heartbeat star in the SMC} \label{sec:heartbeat}

Gaia DR3 4687138015328580736 (OGLE SMC--ECL--5553, Figure \ref{fig:other_panel}g) appears to be an eclipsing heartbeat star in the SMC identified as an eclipsing binary in the Optical Gravitational Lensing Experiment \citep[OGLE]{Udalski92, Pawlak13}.  Heartbeat stars are a rare class of variable stars \citep{Thompson12, Jayasinghe19, Wrona22}, showing tidal distortions near pericenter and tidally-excited oscillations \citep[TEOs,][]{Fuller17}. These stars can be used to empirically study the tidal interactions within binary stars \citep[e.g.,][]{Beck18}. 

OGLE SMC--ECL--5553 has also been identified as an X-ray source in XMM-Newton \citep{Page12} and an infrared variable in the Spitzer SAGE-SMC survey \citep{Gordon11, Polsdofer15}. No TESS light curve is available through the QLP or SPOC pipelines, so we extracted a light curve from the sector 27 full-frame images using {\tt eleanor}. In addition to showing the clear eclipses and brightening between eclipses, the TESS light curves shows additional TEOs with a period of $P_{\rm{TEO}}\simeq1.2$~days. This variability is not seen in the OGLE light curve, which may indicate it comes from a nearby variable star. Since this target is in the SMC, it is outside the footprint of the ALTAS and ZTF variable catalogs but we find no contaminating variables matching the period of the oscillations in the ASAS-SN or OGLE catalogs \citep{Soszynski10, Soszynski11, Pawlak13, Soszynski15, Pawlak16}, so the origin of this variability is unclear.

\subsubsection{Disk Eclipsing Systems} \label{sec:disk_eclipsing}

The two systems in the ``other'' category that have the longest orbital period are Gaia DR3 4655260836825948160 (OGLE LMC--ECL--28271) and Gaia DR3 5884699932561933696 (TT Nor) with periods $P=27.1$~d and $P=37.4$~d, respectively. Both show two eclipses with dramatically different widths and unusual primary eclipse shapes (Figure \ref{fig:other_panel}h and \ref{fig:other_panel}i). In the case of OGLE-LMC-ECL-28271, the primary eclipses last $\sim0.14$ in orbital phase, which is $\sim3.8$~days. In the middle of the primary eclipse, an additional eclipse occurs. The secondary eclipse, separated by $0.5$ in orbital phase, is much narrower and shallower and is more consistent with the additional eclipsing feature seen in the primary eclipse. This system was labeled as an eclipsing binary in OGLE \citep{Pawlak13}, and the OGLE I-band light curve shows the same eclipse features as the ASAS-SN light curve. 

TT Nor also shows two eclipses of extremely different widths and an asymmetric primary eclipse, but here the additional dip in the primary eclipse occurs just before egress. The system is also eccentric, which likely partially contributes to the unequal eclipse widths. Although the orbital period is longer than a TESS sector, two sectors of TESS data are available and provide full orbital phase coverage. Unlike OGLE-LMC-ECL-28271, the secondary eclipse does not resemble the extra `dip' in the primary eclipse. The differences between these two systems could be due to differences in disk size and inclination. 

\section{Discussion}

\begin{table*}
    \centering
    \caption{Summary of group parameters. The ``N TESS'' column reports the number of systems where TESS QLP or SPOC light curves are available. the $P_2$ column reports the number of systems where pulsation/orbital periods could be estimated. The ``N Blend'' column gives the number of binaries where a nearby variable is found in the Gaia, ATLAS, or ZTF catalogs (Table \ref{tab:contamination_table}). The $\texttt{r}1<36\arcsec$ column reports the number of systems where ATLAS reports a distance where the cumulative flux of nearby stars equals the target. The number of targets with X-ray detections are given in the X-ray column, and the MS, SG, RG columns give the evolutionary state of the primary based on the MIST isochrone/evolutionary track divisions.}
    \sisetup{table-auto-round,
         group-digits=false}
    \setlength{\tabcolsep}{3pt}
    \begin{center}
        \begin{tabular}{rrrrrrrrrr}
\toprule
                Group & Total & N TESS & N $P_2$ &     N Blend & $\texttt{r}1<36"$ & N X-ray &  MS &  SG &  RG \\
\midrule
        Spotted Stars &   426 &    398 &     410 & 34 (08.0\%) &       45 (10.6\%) &      68 & 127 & 177 & 105 \\
        Triples/Quads &   225 &    225 &     176 & 54 (24.0\%) &       69 (30.7\%) &       2 & 126 &  33 &  13 \\
        Reg. Pulsator &    36 &     36 &      34 &  5 (13.9\%) &        8 (22.2\%) &       1 &  22 &   4 &   0 \\
           Reflection &    29 &     27 &         &  4 (13.8\%) &        6 (20.7\%) &       2 &  10 &   2 &   3 \\
      Stochastic Var. &    24 &     24 &         &  5 (20.8\%) &        5 (20.8\%) &       0 &   4 &   5 &   1 \\
$\gamma$-Dor Pulsator &    17 &     17 &         &  2 (11.8\%) &        1 (05.9\%) &       0 &  17 &   0 &   0 \\
                Other &     9 &      7 &         &  4 (44.4\%) &        4 (44.4\%) &       1 &   6 &   0 &   0 \\
\bottomrule
\end{tabular}

    \end{center}
    \label{tab:summary_table}
\end{table*}

We present a catalog of eclipsing binaries with additional features in their light curves due to spots, additional eclipsing companions, pulsations, and reflection effects. These binaries were identified through visual inspection of TESS and ASAS-SN light curves in \citetalias{Rowan22}. Table \ref{tab:ep_table} gives the parameters of each system and Table \ref{tab:contamination_table} provides possible blended targets. The full table is available at \url{https://asas-sn.osu.edu/binaries} and in the electronic version of the paper. Table \ref{tab:summary_table} summarizes the results for each group. After grouping the targets by the nature of the extra-physics observed in their light curves, we explore the parameters of each group. 

For spotted stars (Section \S\ref{sec:spotted_stars}), we find the distribution of orbital period is closely tied to evolutionary state (Figure \ref{fig:param_hists}), and few spotted targets are identified on the upper main sequence. Many of these targets are chromospherically active X-ray sources.  We also identify two red $(G_{\rm{BP}} - G_{\rm{RP}} > 1.5$~mag) sub-subgiant candidates. Spectroscopic followup of these systems could be used to determine dynamical masses and radii, which could inform formation and evolutionary pathways \citep{Geller17, Leiner17}.

Our catalog contains \nEE{} targets with extra-eclipses that suggest a hierarchial triple/doubly eclipsing binary configuration (Section \S\ref{sec:extra_eclipses}). However, this group has the largest number of targets with nearby variables and proximity statistic $\texttt{r}1 < 36\arcsec$ (Table \ref{tab:summary_table}), suggesting that some are likely blends of variables due to the low resolution of TESS. Similar to \citet{Zasche19}, we identify an excess of systems with 3:2 period resonance (Figure \ref{fig:p2_result}), which is unlikely to occur from random pairings or blended sources. The detection and characterization of hierarchical triple systems can be used to study dynamical processes such as Kozai cycles and tidal friction that impact the evolution of the period and eccentricity of the inner binary \citep{Mazeh79}. Similar dynamical effects occur for quadruple systems made up of two pairs of binaries \citep{Pejcha13}. For example, CzeV343, which is included in our catalog and was originally identified by \citet{Cagas12}, is a quadruple system with orbital periods in a near 3:2 ratio. \citet{Pejcha22} determined the period and eccentricity of the mutual orbit and confirmed they are bound. The characterization of such systems are relevant to developing models of resonant capture \citep{Tremaine20}.

We perform a similar period calculation procedure for the regular pulsators and $\gamma$-Dor systems (Sections \S\ref{sec:regular_pulsators} and \S\ref{sec:gamma_dor_pulsators}), identifying multiple significant frequencies for the later group. Most, if not all, of the Cepheid and RR Lyrae pulsators appear to be blends, highlighting the need for additional photometry to confirm these systems. Systems with $\delta$-Scuti, $\gamma$-Dor, and $\beta$-Cephei pulsations may be used to place additional constraints on stellar parameters and improve understanding of stellar interiors \citep[e.g.,][]{Johnston19}. We also find \nSto{} binaries with stochastic low frequency variability (Section \S\ref{sec:stochasic_variability}), which likely arises from gravity waves in massive stars \citep{Bowman19}. Spectroscopic followup of these systems could be used to derive dynamical masses and make direct comparisons with asteroseismic models, improving our understanding of stellar interiors. 

Finally, we investigate \nOther{} systems that don't fit into these groups (Section \S\ref{sec:other_targets}). We highlight two eclipsing CVs, two disk systems, a heartbeat star, and an eclipsing subdwarf. Four of these systems could be considered possible blends based on crossmatches to variable star catalogs or the $\texttt{r1}$ proximity statistic (Table \ref{tab:summary_table}). We identify two systems, Gaia DR3 899052367257785984 (Section \S\ref{sec:ellipsoidal_triple}) and Gaia DR3 6089677590337973760 (Section \S\ref{sec:ee_gammador}) as clear blends. EC~21178--5417 is unlikely to be a blend given given the CMD position and previous detailed studies. The disk eclipsing variability seen in OGLE LMC--ECL--28271 is observed by both ASAS-SN and OGLE, making this system also unlikely to be a blend. Characterization of these exotic systems can be used to study unique stages of binary evolution \citep[e.g.,][]{Ostensen10, Khangale20}.

It is clear, however, that the low resolution of TESS makes the blending of distinct variables a considerable problem. Additional photometric and spectroscopic observations of these targets can be used to confirm the nature of the variability and derive dynamical masses and radii. All of the systems reported in this catalog were identified as eclipsing binaries in ASAS-SN \citet{Jayasinghe21}, and many of the ASAS-SN light curves are clear outliers from standard eclipsing binaries. The complementary nature of the long-baseline, daily cadence ASAS-SN light curves and the short duration, high precision TESS photometry allows for accurate identification and characterization of the ``extra-physics'' in these systems.

\section*{Acknowledgements}
We thank JJ Hermes for his valuable comments and discussion. We thank Las Cumbres Observatory and its staff for their continued support of ASAS-SN. ASAS-SN is funded in part by the Gordon and Betty Moore Foundation through grants GBMF5490 and GBMF10501 to the Ohio State University, and also funded in part by the Alfred P. Sloan Foundation grant G-2021-14192.
 
DMR, TJ, KZS and CSK are supported by NSF grants AST-1814440 and AST-1908570. Support for TJ was provided by NASA through the NASA Hubble Fellowship grant HF2-51509 awarded by the Space Telescope Science Institute, which is operated by the Association of Universities for Research in Astronomy, Inc., for NASA, under contract NAS5-26555.

This work has made use of data from the European Space Agency (ESA)
mission {\it Gaia} (\url{https://www.cosmos.esa.int/gaia}), processed by
the {\it Gaia} Data Processing and Analysis Consortium. 

This paper includes data collected with the \textit{TESS} mission, obtained from the MAST data archive at the Space Telescope Science Institute (STScI). Funding for the TESS mission is provided by the NASA Explorer Program. STScI is operated by the Association of Universities for Research in Astronomy, Inc., under NASA contract NAS 5-26555. CSK, KZS and DMR TESS research is supported by NASA grant 80NSSC22K0128.

\section*{Data Availability}

The ASAS-SN photometric data underlying this article are available in the ASAS-SN eclipsing
binaries database (https://asas-sn.osu.edu/binaries) and the ASAS-SN Photometry Database
(https://asas-sn.osu.edu/photometry). 
The data underlying this article are available in the article and in its online supplementary material.



\bibliographystyle{mnras}
\bibliography{extraphysics, heasarc_master} 




\appendix

\section{Light Curves of Selected Targets} \label{sec:appendix}

\begin{figure*}
    \centering
    \includegraphics[width=\linewidth]{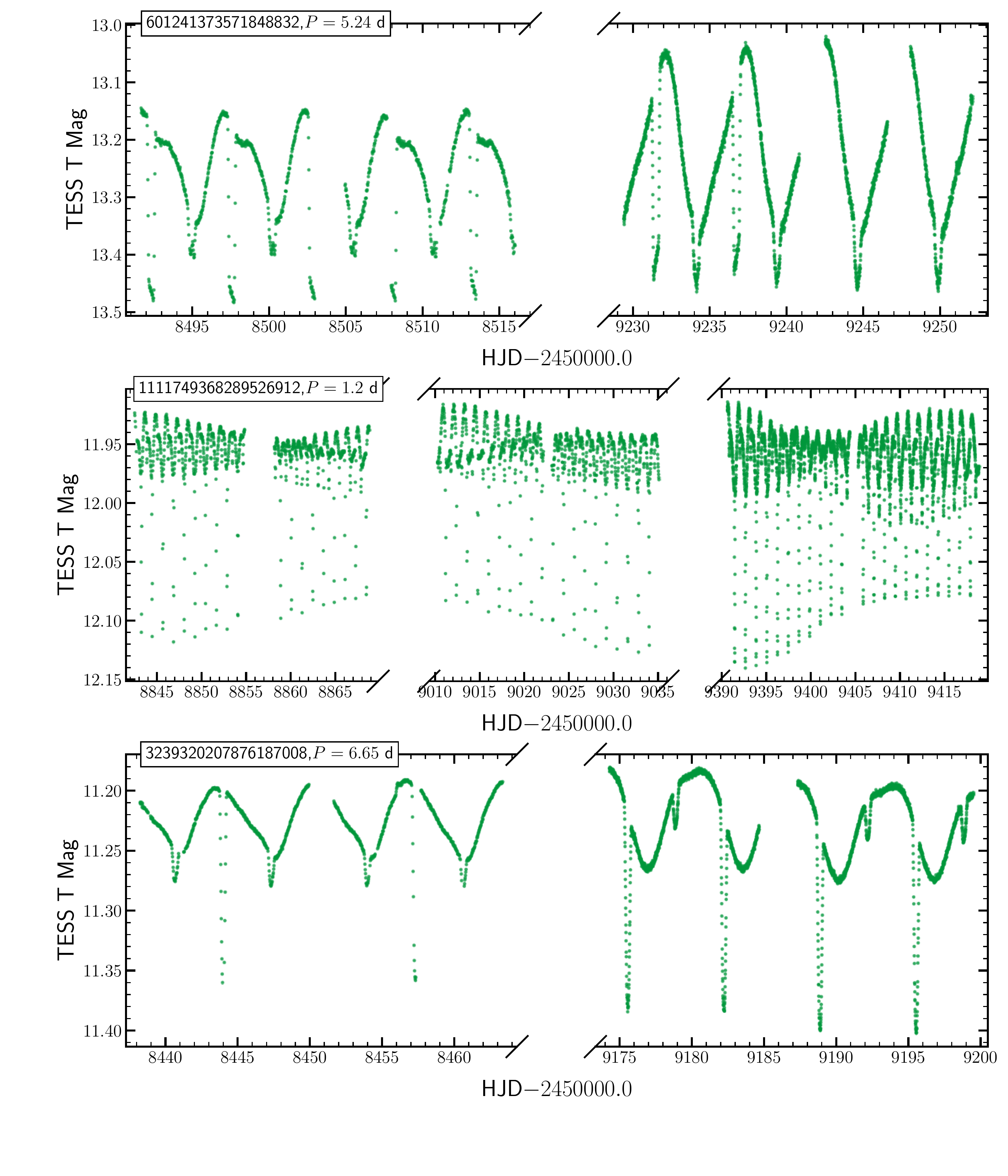}
    \caption{Three examples of spotted eclipsing binaries where there is clear rotational modulation in the TESS light curves. The orbital period and Gaia DR3 Source ID are given in the upper left of each panel.}
    \label{fig:appendix_spots}
\end{figure*}

\begin{figure*}
    \centering
    \includegraphics[width=\linewidth]{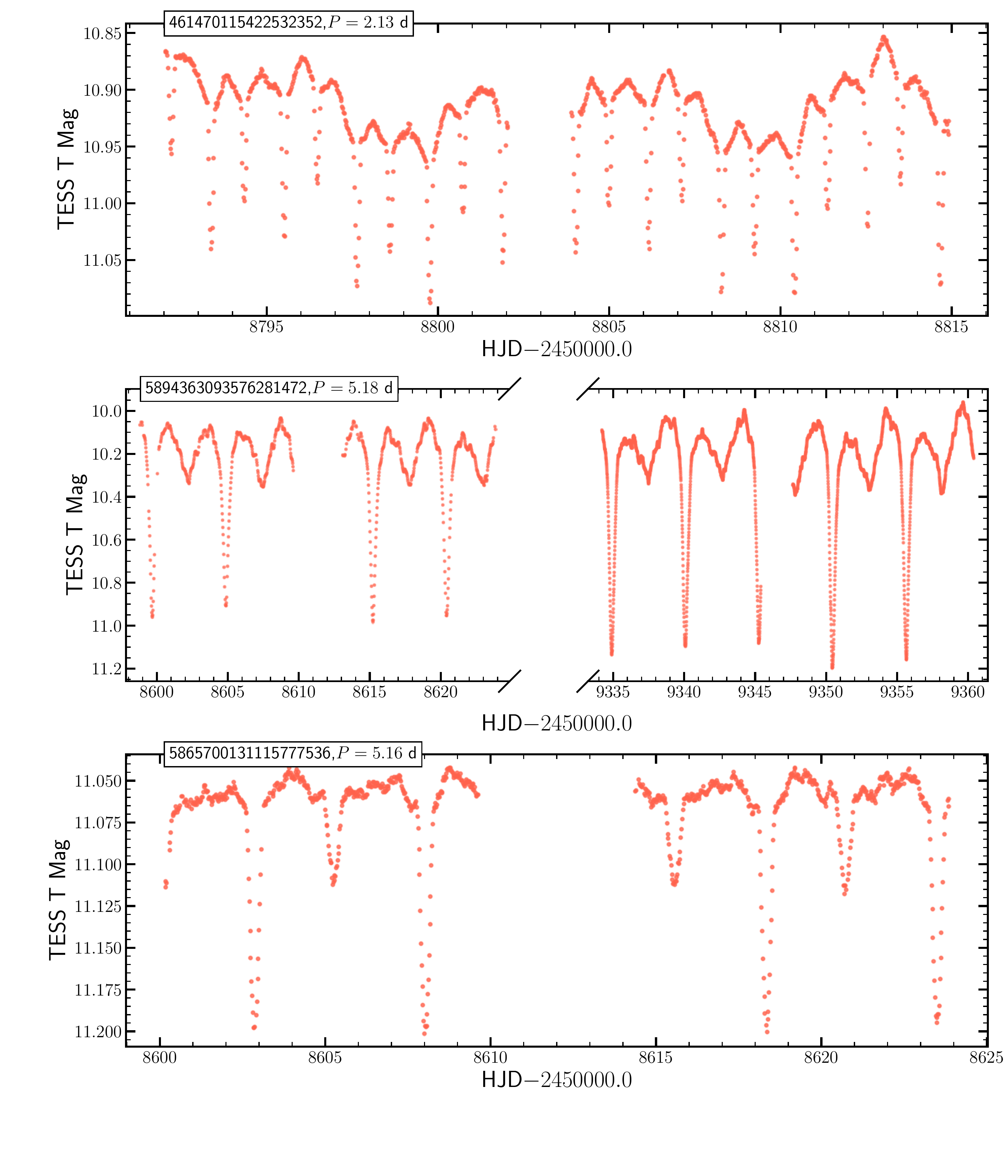}
    \caption{Three examples of TESS light curves for stochastically varying eclipsing binaries. The orbital period and Gaia DR3 Source ID are given in the upper left of each panel.}
    \label{fig:appendix_sto_var}
\end{figure*}

\begin{figure*}
    \centering
    \includegraphics[width=\linewidth]{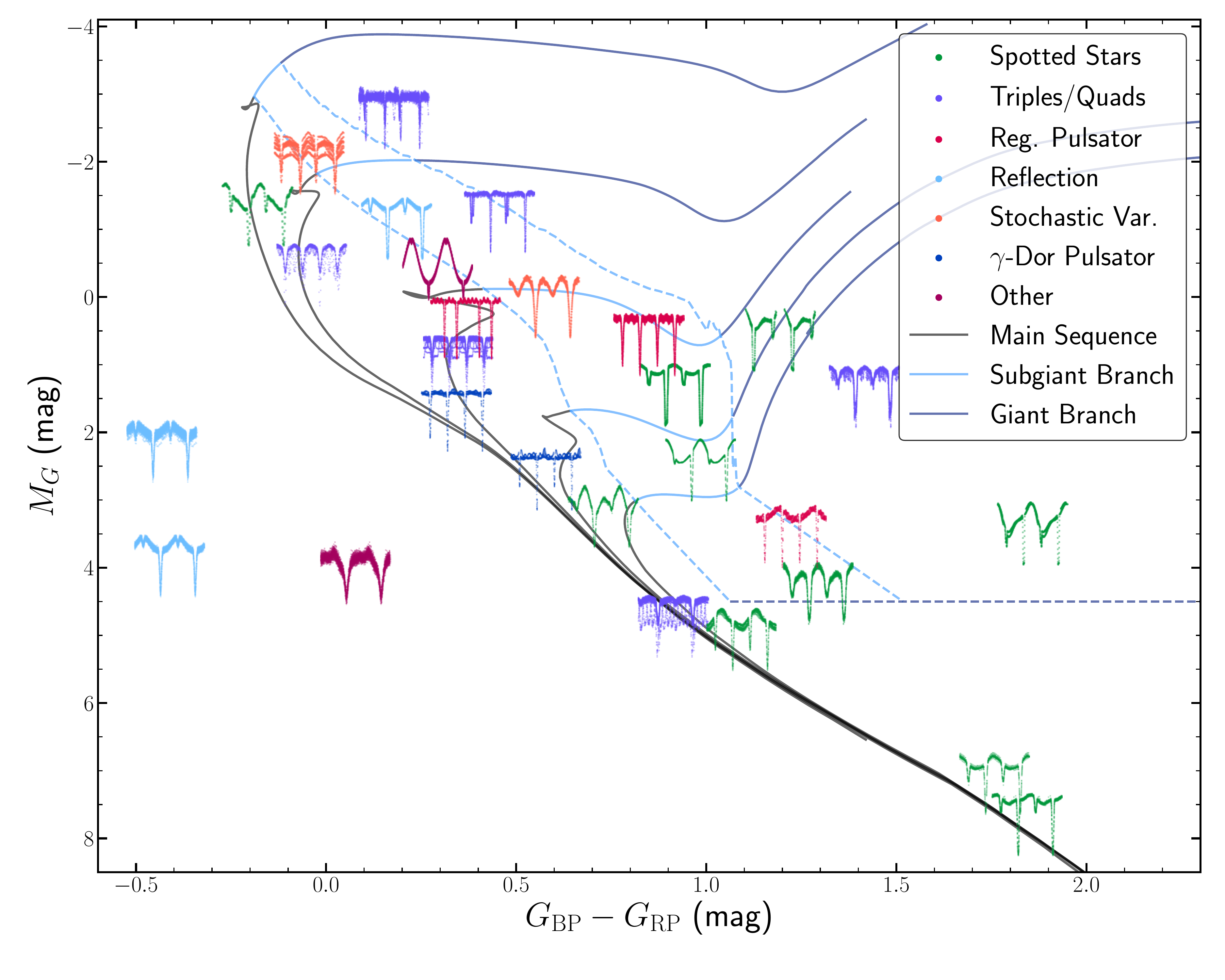}
    \caption{Examples of eclipsing binaries of different extra-physics types on a Gaia CMD.}
    \label{fig:appendix_cmd}
\end{figure*}

All light curves from Table \ref{tab:ep_table} are available at \url{https://asas-sn.osu.edu/binaries}. Here we show example light curves to highlight different types of variability. Figure \ref{fig:appendix_spots} shows additional examples of rotational variability where the spot behavior changes throughout and between TESS sectors. Figure \ref{fig:appendix_sto_var} shows three examples of binaries with stochastic variability. Finally, Figure \ref{fig:appendix_cmd} shows additional examples on a Gaia CMD. 

\bsp	
\label{lastpage}
\end{document}